\documentclass{aa}   

\usepackage{natbib}
\bibpunct{(}{)}{;}{a}{}{,} % to follow the A&A style

\usepackage{graphicx}
\usepackage{txfonts}
\usepackage[normalem]{ulem}
\usepackage{xcolor}
\usepackage{amsmath}
\usepackage{amssymb}
\usepackage{bm}
\usepackage{subcaption}
\usepackage{multirow}

\usepackage{cancel}

\begin{document}

   \title{Mitigating baryonic effects in weak lensing with higher-order statistics}

   \author{%
    Andreas Tersenov \inst{1,2,3}\fnmsep\thanks{\email{atersenov@physics.uoc.gr}}
    \and
    Sacha Guerrini \inst{4}
    \and
    Jean-Luc Starck
    \inst{2,3}
    \and
    Martin Kilbinger
    \inst{3}
    }

   \institute{
            Department of Physics, University of Crete, Greece
            \and
             Institutes of Computer Science and Astrophysics, Foundation for Research and Technology Hellas (FORTH), Greece
             \and
             Université Paris-Saclay, Université Paris Cité, CEA, CNRS, AIM, 91191, Gif-sur-Yvette, France
             \and
             Universit\'e Paris Cit\'e, Universit\'e Paris-Saclay, CEA, CNRS, AIM, F-91191, Gif-sur-Yvette, France
   }

   \date{Received; accepted}

  \abstract
  % context heading (optional)
  % {} leave it empty if necessary  
   {
    Weak gravitational lensing is a premier cosmological probe, but its small-scale statistical power is compromised by baryonic feedback. While higher-order statistics (HOS) capture essential non-Gaussian information, their sensitivity to these astrophysical processes remains a critical concern for upcoming Stage IV surveys.
   }
  % aims heading (mandatory)
   {
    We investigate the information content and robustness of the angular power spectrum (PS), starlet peak counts, and the starlet $\ell_1$-norm in the presence of unmodeled baryonic feedback. We determine the necessary scale cuts to mitigate parameter biases across various survey areas and evaluate whether HOS provide tighter constraints than the PS on these ``baryon-safe'' scales. Finally, we test the Bernardeau-Nishimichi-Taruya (BNT) transform as an alternative mitigation strategy.
   }
  % methods heading (mandatory)
   {
    We employ the cosmoGRID V1 simulation suite, which uses a baryon correction model to imprint feedback signatures on dark matter convergence maps. We derive cosmological constraints via simulation-based inference, using neural posterior estimation with masked autoregressive flows across footprints ranging from Stage III-like to the full sky.
   }
  % results heading (mandatory)
   {
    Biases from unmodeled baryons grow with survey area, exceeding $2\sigma$ for all three statistics at Stage IV-like footprints.
    Recovering unbiased parameters requires survey-area-dependent scale cuts that discard a substantial fraction of the signal. Restricted to these baryon-safe scales, the starlet $\ell_1$-norm still reaches a figure of merit almost twice that of the PS. The BNT transform localizes the baryonic sensitivity to the lowest transformed redshift bin and improves the PS figure of merit by a factor of $\sim$1.4, while its linear mixing of shape noise inflates the posterior contours of the map-based HOS.
   }
   {
    Non-Gaussian information survives even the most conservative treatment of baryonic feedback. Higher-order statistics therefore deliver a substantial gain over the power spectrum with no baryonic modeling at all, and an even larger one as modeling improves.
   }


   \maketitle
%
%-------------------------------------------------------------------

\section{Introduction} \label{sec:intro}

Weak gravitational lensing (WL), the coherent distortion of distant galaxy images by the intervening large-scale structure, has established itself as a premier probe of the matter density field \citep{bartelmann_weak_2001, kilbinger_cosmology_2015, mandelbaum_weak_2018}. By tracing the distribution of dark matter, WL provides a unique window into the growth of structure and the nature of cosmic acceleration. The current generation of photometric surveys has already placed tight constraints on the standard $\Lambda$CDM cosmological model. However, the field is transitioning into the era of Stage IV surveys, including the Euclid mission \citep{euclid2024}, the Vera C. Rubin Observatory with the Legacy Survey of Space and Time (LSST) \citep{ivezic_lsst_2019}, and the Roman Space Telescope \citep{spergel_roman_2015}. These missions will cover vast sky areas with unprecedented depth, promising an order-of-magnitude leap in constraining power.

To realize the full potential of Stage IV data, we must extract cosmological information from the non-linear scales of the cosmic web. It is on these small angular scales where the bulk of the WL signal resides, but also where the matter field becomes highly non-Gaussian \citep{weinberg_observational_2013}. Traditional analyses, relying primarily on the angular power spectrum (PS) or two-point statistics, capture the full information content only if the underlying field is Gaussian. Consequently, reliance on 2-point statistics alone is fundamentally incomplete in the non-linear regime. Capturing the information they miss requires statistics that probe these non-Gaussian features.

This necessity has driven the development of various higher-order statistics (HOS), including the bispectrum \citep{Takada_three-point_2003, takada_cosmological_2004}, Minkowski functionals \citep{kratochvil_probing_2012,grewal_minkowski_2022}, and higher-order moments (skewness, kurtosis) of the convergence field. Statistics based on local extrema, such as peak counts \citep{Marian_cosmology_2009, dietrich_cosmology_2010, lin2015new, aycoberry_unions_2023}, have proven particularly effective, tracing the halo mass function and breaking parameter degeneracies inherent in 2-point analyses. More recently, full-field summary statistics such as the scattering transform \citep{cheng_new_2020, valogiannis_towards_2022} or the wavelet $\ell_1$-norm have emerged as powerful tools to capture information from the entire map, moving beyond simple feature counting.

However, the transition to non-linear scales introduces a major systematic challenge: baryonic feedback. Astrophysical processes such as supernovae and active galactic nuclei (AGN) feedback significantly alter the matter distribution, redistributing gas and suppressing the matter PS on small to intermediate scales \citep{van_daalen_effects_2011, semboloni_quantifying_2011, chisari_modelling_2019}. If unmodeled, this suppression mimics the suppression of structure growth caused by massive neutrinos or dark energy, introducing biases in cosmological parameters.

Standard mitigation strategies typically fall into two categories: explicit modeling of the feedback mechanisms or the application of conservative scale cuts \citep{eifler_accounting_2015, mead_accurate_2015}. Considerable effort has gone into the modeling route, which now extends from two- and three-point shear statistics \citep{semboloni2013effect, arico2021simultaneous} to map-level prescriptions that reproduce weak lensing higher-order statistics \citep{anbajagane_map-level_2024, zhou_map-level_2025}, and it is this progress that allows analyses to retain smaller scales. Modeling approaches, whether relying on hydrodynamical prescriptions, halo models, or approximate parametric forms such as baryon correction models \citep{schneider_new_2015, schneider2019quantifying}, are nevertheless limited by our incomplete understanding of sub-grid physics. Different hydrodynamical simulations often yield divergent predictions for the magnitude and shape of the matter PS suppression, leading to significant model uncertainty \citep{chisari_modelling_2019, villaescusa2021camels}, and a prescription calibrated on the power spectrum does not automatically reproduce higher-order statistics \citep{lee_impact_2026}. Relying on an inaccurate feedback model can therefore introduce biases in the inferred cosmological parameters that are difficult to quantify. Conversely, applying conservative scale cuts to remove contaminated data is robust against these modeling errors but inevitably discards the high-signal-to-noise information residing on small scales. Whichever route is taken, the scales that carry the most constraining power are the ones where theoretical predictions are least reliable.

This tension motivates the search for estimators that maximize information extraction on the intermediate scales where baryonic physics is sub-dominant. HOS can be measured at any angular scale, and current surveys already measure them well into the quasi-linear regime, in DES \citep{zurcher2022dark, gatti2022dark, gatti2025dark, gomes2025dark, prat2026dark}, KiDS \citep{heydenreich2021persistent, burger2023kids, harnois-deraps_kids1000_2024}, HSC \citep{cheng_cosmological_2024, armijo2025cosmological, novaes2025cosmology, sugiyama2025cosmology}, and UNIONS \citep{aycoberry_unions_2023}, with the same statistics being prepared for \textit{Euclid} \citep{ajani_forecasts_2023, vinciguerra_euclid_2026}. They are nonetheless most commonly associated with the deeply non-linear regime, where the departures from Gaussianity are largest, and that is also where baryonic feedback is strongest. What they offer once the contaminated scales are removed is therefore a separate question. Where those scales lie is itself set by the statistical precision of the survey, so a cut derived for a Stage III footprint need not hold at Stage IV. How strongly the feedback affects them in the first place is also unevenly characterized. Most assessments compare summary statistics measured in hydrodynamical and dark matter-only simulations \citep[e.g.,][]{broxterman_flamingo_2023, lee2023comparing}, and fewer carry the difference through to the inferred parameters \citep[e.g.,][]{grandon_impact_2024}. Several of the statistics now used on data have not been tested against baryons at all, among them the starlet $\ell_1$-norm.

In addition, lensing geometry makes any scale cut more costly than it needs to be. The lensing kernels are broad, so each tomographic bin collects structure over a wide range of distances, and a fixed angular scale mixes the small physical scales of nearby lenses with the much larger ones of distant lenses. A cut in angular scale cannot separate the two, so removing the low-redshift structures where baryonic feedback acts also discards clean information from lenses at high redshift. Data transformations such as the Bernardeau-Nishimichi-Taruya \citep[BNT;][]{bernardeau_cosmic_2014} transform are designed to break this coupling by nulling the sensitivity to small-scale low-redshift lenses, which could mitigate baryonic uncertainties without simply discarding data. Its benefit has been explored for two-point statistics \citep{taylor2018kcut, taylor2021xcut, gu2025mitigating, gu2026mitigating}, whereas the behavior of higher-order statistics is less well investigated, especially under full posterior inference.

In this work we carry the effect of unmodeled baryons through to the inferred cosmological parameters, for the PS, starlet peak counts, and the starlet $\ell_1$-norm. We do this with the \texttt{cosmoGRID} simulation suite \citep{fluri_full_2022, kacprzak_cosmogridv1_2023} and a simulation-based inference (SBI) framework, which avoids assuming a form for the likelihood of the non-Gaussian statistics. Specifically, we aim to:
\begin{enumerate} 
    \item Quantify the bias induced by baryons as a function of survey area, identifying the tipping point for Stage IV surveys. 
    \item Determine the necessary scale cuts for each statistic to ensure unbiased inference. 
    \item Compare the constraining power of the HOS and the PS once each is restricted to its own baryon-safe scales.
    \item Evaluate the BNT transform as a mitigation strategy, for the PS and for the map-based HOS.
\end{enumerate}

The paper is structured as follows: Section \ref{sec:simulations} details the \texttt{cosmoGRID} N-body simulation suite and the modeling of baryonic feedback. Section \ref{sec:stats} defines the weak lensing summary statistics utilized in our analysis (the angular power spectrum, starlet peak counts, and the starlet $\ell_1$-norm) alongside the theoretical framework of the BNT transform. Section \ref{sec:inference} outlines our simulation-based inference pipeline using neural posterior estimation, and the experimental setup used to vary the survey area. Section \ref{sec:results} presents our main inference results, including the area-dependent scaling of baryonic bias, the determination of robust scale cuts, the information content preserved on quasi-linear scales, and the performance of the BNT transform. Finally, we summarize our conclusions in Section \ref{sec:conclusions}. The appendices collect the definitions of the summary statistics (Appendix \ref{app:stats}), the measured multipole coverage of the starlet bands (Appendix \ref{app:starlet_ell}), and supporting figures for the noiseless measurements, the peak counts, and the BNT transform (Appendices \ref{app:noiseless_stats}--\ref{app:bnt_viz}).

\section{Simulations} \label{sec:simulations}
\subsection{The \texttt{cosmoGRID V1} simulation suite} \label{sec:cosmoGRID}

To perform our analysis, we use the \texttt{cosmoGRID V1} simulation suite \citep{fluri_full_2022, kacprzak_cosmogridv1_2023}, a large-scale lightcone dataset designed specifically for map-level analysis of large-scale structure probes. This suite was generated using the \texttt{PkdGrav3} N-body code \citep{sgier_fast_2019}, and is optimized for parameter measurement necessary for Stage III surveys, including KiDS, DES, and HSC.

The simulations span a flat $w\mathrm{CDM}$ cosmological model, varying six parameters: the total matter density ($\Omega_\textrm{m}$), the amplitude of matter fluctuations ($\sigma_8$), the dark energy equation-of-state parameter ($w_0$), the Hubble constant ($H_0$), the scalar spectral index ($n_\textrm{s}$), and the baryon density ($\Omega_\textrm{b}$). The sum of neutrino masses is fixed at ($\Sigma m_\nu=0.06$) eV.

To densely and efficiently cover this parameter space, the cosmologies were sampled uniformly using a 6-dimensional Sobol sequence. The suite is divided into two sub-grids of 1,250 points each: a ``wide'' prior set encompassing broad, observationally uninformative boundaries, and a ``narrow'' prior set tightly focused around the fiducial cosmology. The bounds for the wide uniform priors are summarized in Table \ref{tab:cosmogrid_priors}.

In addition to these bounds, two specific regions of the parameter space were excluded. First, a restriction in the $\Omega_m - w_0$ plane removes models where $w_0 < -1.1$ at certain matter densities. Second, a cut is applied in the $\Omega_m - \sigma_8$ plane to restrict the sample to the observationally motivated $S_8$ degeneracy region. We refer the reader to \citet{fluri_full_2022, kacprzak_cosmogridv1_2023} for exhaustive details regarding the prior design and sampling methodology.

\begin{table}[t]
\centering
\begin{tabular}{lc}
\hline \hline
\textbf{Cosmological parameter}  & \textbf{Prior range} \\ \hline
 $\Omega_m$ & [0.10,0.50] \\
 $\sigma_8$ & [0.40,1.40] \\
 $w_0$  & [-2.00,-0.33] \\
$H_0$ & [64.0,82.0] \\
 $n_s$ & [0.87,1.07] \\
 $\Omega_b$ & [0.03,0.06] \\
\hline
\end{tabular}
\caption{Cosmological parameters varied in the \texttt{cosmoGRID V1} simulation suite and their respective broad uniform prior distributions \citep{fluri_full_2022}.}
\label{tab:cosmogrid_priors}
\end{table}

The core dataset contains 2,500 distinct cosmologies, each with 7 independent N-body realizations to provide sufficient volume for training deep learning models. The suite also includes 200 independent fiducial realizations \footnote{The fiducial setting corresponds to $\Omega_m=0.26, \sigma_8 = 0.84, b = 0.0493, n_s = 0.9649, H_0 = 67.36, w =-1.0, A_{\rm IA} = 0.0$, and $\eta = 0.0$.} and a benchmark set for validation.

\begin{figure}[t]
    \centering
    \includegraphics[width=0.48\textwidth]{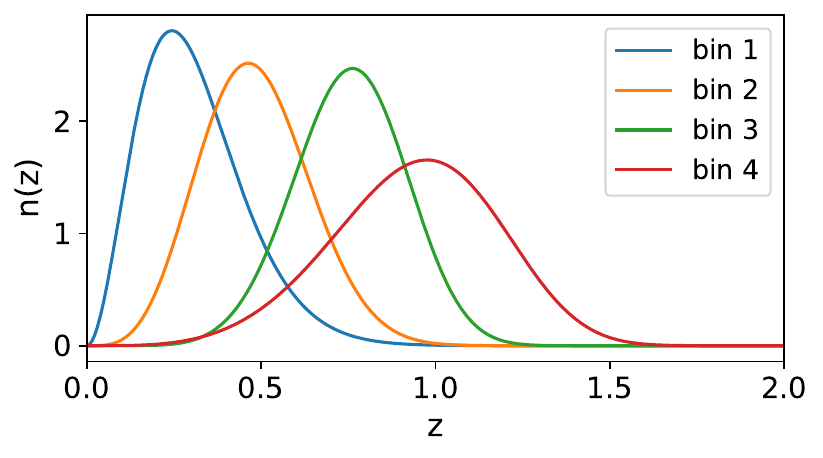}
    \caption{
    \textbf{Source redshift distributions.} The normalized galaxy number density $n_i(z)$ for the four tomographic bins used in this analysis. The bins correspond to mean redshifts of $z \approx [0.3,0.5,0.7,0.9]$.
    }
    \label{fig:nz_bins}
\end{figure}

For this analysis, we use the ``Stage-III forecast probe maps'' subset. These are full-sky projected weak lensing convergence maps provided at a HEALPix \citep{gorski_healpix_2005} resolution of $N_{\rm side}=512$. This resolution is sufficient to probe the angular scales of interest ($\ell \leq 1024$) while maintaining a manageable data volume for the large ensemble required by SBI.
The maps are provided in four tomographic redshift bins, chosen to mimic the sensitivity of a typical Stage III weak lensing survey (mean redshifts $z \approx [0.3, 0.5, 0.7, 0.9]$). The normalized redshift distributions $n_i(z)$ for these $z$-bins are shown in Figure \ref{fig:nz_bins}. 

Importantly for this work, the \texttt{cosmoGRID} suite provides paired dark matter-only (DMO) and ``baryonified'' maps for every cosmology. The latter are generated using a map-level implementation of the baryon correction model (BCM), which modifies the density field to mimic the effects of stellar and AGN feedback, as detailed in the following section. 

To robustly capture cosmic variance during inference, we use the suite's shell permutation scheme to generate multiple quasi-independent map realizations for each cosmology. To emulate realistic Stage IV observational conditions, we add uncorrelated Gaussian shape noise to these mock convergence maps. The noise variance in each pixel is computed assuming a Euclid-like survey configuration, defined by an expected effective galaxy number density of $n_{\rm gal}=30\,\textrm{arcmin}^{-2}$ (distributed across our four tomographic bins) and an intrinsic ellipticity dispersion of $\sigma_\epsilon=0.26$ per component.

\subsection{Modeling baryonic effects} \label{sec:sim_baryonic_effects}

To account for the impact of non-gravitational physics, which is the dominant systematic uncertainty on small scales, the suite incorporates baryonic feedback via the BCM \citep{schneider_new_2015, schneider2019quantifying}. The BCM is a phenomenological framework designed to modify the density field of DMO simulations to reproduce the structural changes observed in high-resolution hydrodynamical simulations.

The model operates by redistributing mass within dark matter halos via a flexible displacement field, mapping the initial Navarro-Frenk-White (NFW) density profile ($\rho_{\rm nfw}$) into a baryon-corrected composite profile ($\rho_{\rm bcm}$). This final profile comprises a compact central stellar component, bound hot gas ejected by feedback, and an adiabatically relaxed dark matter halo. To bypass the prohibitive computational cost of traditional 3D particle displacement for massive simulation-based inference datasets, the suite employs the ``shell baryonification'' technique \citep{fluri_full_2022, kacprzak_cosmogridv1_2023}. By applying the BCM displacement corrections directly to the pixels of projected 2D density shells (lens planes) rather than 3D particles, this approach achieves a near 1,000-fold speed-up, enabling consistent baryonic modeling across the entire grid of 2,500 cosmologies.

The strength of the correction is controlled by a characteristic halo mass $M_c$, below which halos are increasingly depleted of gas, with a power-law redshift evolution $M_c = M_c^0 (1+z)^\nu$. The fiducial baryonified maps used here adopt $\log_{10}(M_c^0 / h^{-1}M_\odot) = 13.82$ and $\nu = 0$, the remaining parameters being fixed to values calibrated on X-ray gas fractions and the hydrostatic mass bias \citep{kacprzak_cosmogridv1_2023}.

\section{Weak lensing summary statistics} \label{sec:stats}

The robust quantification of the WL signal relies on summary statistics that compress the high-dimensional data (pixelised convergence maps) into a tractable, yet informative, format. 
We employ three different statistics, 
establishing the classical two-point measure as our baseline before exploring higher-order metrics:
\begin{itemize}
    \item PS ($C_\ell$): The standard metric that measures the two-point angular correlation of the convergence field as a function of multipole $\ell$. While it captures all information for a purely Gaussian random field, it misses the non-linear features dominant on small scales.
    \item Starlet peak counts \citep{ajani_constraining_2020}: A non-Gaussian statistic that counts local maxima in the convergence field. To separate physical scales, we apply the isotropic undecimated wavelet (``starlet'') transform. Peaks in these wavelet coefficient maps are sensitive tracers of massive dark matter halos and structure alignment.
    \item Starlet $\ell_1$-norm \citep{ajani_starlet_2021}: A multiscale characterization of the full probability density function of the convergence field. By summing the absolute values of the starlet wavelet coefficients within specific signal-to-noise ratio (SNR), $\nu$, bins, it captures the morphology of both overdensities (filaments/halos) and underdensities (voids) without the ambiguity of defining discrete features.
\end{itemize}

Detailed mathematical definitions for each statistic, including the formal multiscale decomposition for the starlet-based statistics, are provided in Appendix \ref{app:stats}. In the following subsections, we focus directly on quantifying how unmodeled baryonic feedback impacts these summary statistics across different physical scales and SNR regimes. The fractional differences shown in this section are measured on full-sky maps. This uses the full simulation volume and makes the shaded bands the smallest statistical uncertainty any survey could achieve. Survey area enters the analysis in Sect. \ref{sec:results}, which uses masked footprints.

\subsection{Angular power spectrum (PS)}

\begin{figure}[t]
    \centering
    \includegraphics[width=0.48\textwidth]{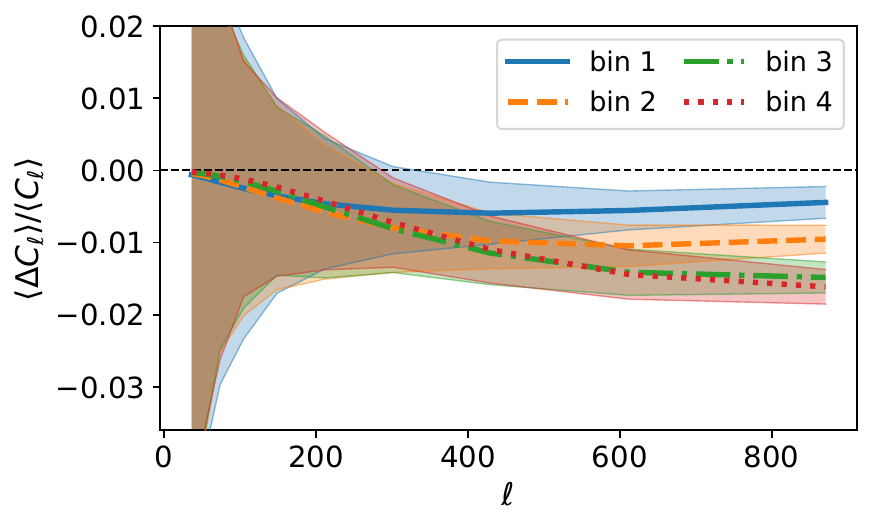}
    \caption{
    Fractional difference between the auto-components of the PS measured on baryonified versus DMO full-sky maps across the four tomographic redshift bins, with matched shape-noise realizations between each pair of baryonified and DMO maps. Solid lines indicate the mean relative difference over the fiducial realizations, and shaded bands show the $1\sigma$ statistical uncertainty of the measured spectrum, $\mathrm{std}(C_\ell^{\mathrm{bar}})/\langle C_\ell^{\mathrm{dmo}}\rangle$.
    }
    \label{fig:frac_diff_ps}
\end{figure}

In this work, we compute the power spectra numerically from the HEALPix convergence maps using the \texttt{NaMaster} package\footnote{\url{https://github.com/LSSTDESC/NaMaster}} \citep{alonso_unified_2019}.
Figure \ref{fig:frac_diff_ps} illustrates the impact of the BCM on the auto-components of the PS for the fiducial cosmology. We plot the fractional difference between the mean baryonified spectrum and the mean DMO spectrum: 
\begin{equation} 
    \frac{\Delta C_\ell}{C_\ell} = \frac{\langle C_\ell^{\mathrm{bar}} - C_\ell^{\mathrm{dmo}} \rangle}{\langle C_\ell^{\mathrm{dmo}} \rangle}. 
\end{equation} 
Each baryonified map shares the shape-noise realization of its DMO counterpart, so noise and cosmic variance cancel in the difference. The shaded regions show the $1\sigma$ statistical uncertainty of the measured spectrum, $\mathrm{std}(C_\ell^{\mathrm{bar}})/\langle C_\ell^{\mathrm{dmo}}\rangle$.

We observe two primary trends in this fractional difference. First, regarding scale dependence, the deviation increases with multipole $\ell$ as expected. On large scales ($\ell \leq 300$), the physics is dominated by linear gravity, and the impact of baryons is negligible. At smaller scales ($\ell > 500$), the feedback mechanisms in the BCM significantly suppress the matter density, leading to an apparent deviation that reaches $\approx 1.5\%$ by $\ell \approx 1000$ for the higher-redshift bins.

Second, we note a strong redshift dependence. In our measurements, the fractional baryonic bias appears most prominent in the higher-redshift source bins (bins 3 and 4). However, this apparent trend is an artifact of the SNR imposed by galaxy shape noise \citep{yang_baryon_2013, martinet_probing_2021}. When computing the fractional difference on noisy maps, the shape noise power $N_\ell$ contributes to the denominator: $\Delta C_\ell / (C_\ell^{\mathrm{dmo}} + N_\ell)$. Because the intrinsic cosmological signal $C_\ell^{\mathrm{dmo}}$ is substantially weaker for low-redshift sources, the constant noise term $N_\ell$ rapidly dominates the denominator at smaller scales. This artificially drives the fractional difference toward zero, washing out the true baryonic bias within the statistical variance. Conversely, the stronger intrinsic signal of high-redshift sources ensures the denominator remains signal-dominated to higher multipoles, allowing the baryonic suppression to remain visible above the noise floor.

We explicitly verify this in Appendix \ref{app:noiseless_stats} (Fig. \ref{fig:frac_diff_ps_noiseless}), where we present the same measurements extracted from noiseless maps. In the absence of shape noise, the recovered physical behavior aligns exactly with theoretical expectations from hydrodynamical simulations and analytical models \citep{mohammed_baryonic_2014, huang2019modelling}: the lowest redshift bin is the most heavily suppressed by baryons, and its deviation from the DMO prediction begins at larger angular scales (lower $\ell$) than in the higher-redshift bins \citep[e.g.,][]{chisari2018impact, schneider2019quantifying, van2020exploring, mead_accurate_2015}.

\subsection{Higher-order statistics} \label{sec:hos}

For the starlet peak counts and the starlet $\ell_1$-norm, we decompose the convergence maps into four dyadic wavelet scales and one coarse scale using the starlet transform \citep{starck:sta06}. The starlet acts as a band-pass filter. Following convention, we refer to the first four wavelet bands by the angular scales $[10\arcmin, 20\arcmin, 40\arcmin, 80\arcmin]$. Our data vectors use these four wavelet scales, with the coarse scale excluded. The measured multipole coverage of each band is presented in Appendix~\ref{app:starlet_ell}.

To extract the summary statistics, we define distinct binning schemes for each method. For the starlet peak counts, we compute the histogram of local maxima $N_{\rm peaks}(\nu)$ using 40 linearly spaced bins between $\nu_{\min} = -2$ and $\nu_{\max} = 10$. This histogram extends to negative values to capture weak local high-points embedded within globally underdense environments. 
In contrast, for the starlet $\ell_1$-norm, we compute the sum of the absolute values of the wavelet coefficients using 40 linearly spaced bins between $\nu_{\min} = -10$ and $\nu_{\max} = 10$. This range extends much deeper into the negative regime than the starlet peak counts because the $\ell_1$-norm samples the entire field, requiring a wider range to properly capture the deep under-densities (voids) that local maxima rarely reach. 
For both methods, the binning is applied independently to each wavelet scale $j$, and the resulting vectors are concatenated to form the final summary statistic. By separating the signal into distinct frequency bands, we can isolate the impact of baryonic feedback—which predominantly affects the finest scales—while preserving the clean cosmological signal in the larger wavelet scales.

\begin{figure*}[t]
    \centering
    \includegraphics[width=0.95\textwidth]{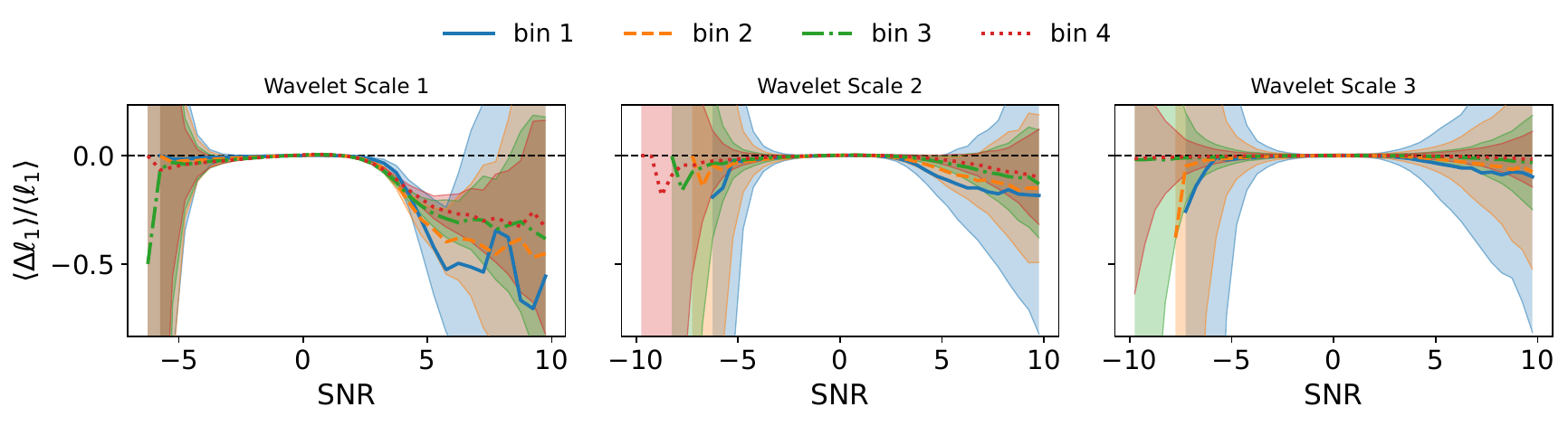}
    \caption{
        Fractional difference in the starlet $\ell_1$-norm between baryonified and DMO full-sky maps as a function of SNR ($\nu$), with matched shape-noise realizations between each pair of baryonified and DMO maps. The panels display results for the first three wavelet scales ($j=1,2,3$, left to right). Lines indicate the mean fractional difference over the fiducial realizations, and shaded bands show the $1\sigma$ statistical uncertainty of the measured $\ell_1$-norm. Bins in which the DMO statistic vanishes are masked.
    }
    \label{fig:frac_diff_l1}
\end{figure*}

Figure \ref{fig:frac_diff_l1} quantifies the bias induced by baryons on the starlet $\ell_1$-norm across the first three wavelet scales. Because the starlet peak counts exhibit a highly similar qualitative response to the BCM as the starlet $\ell_1$-norm, we present the corresponding fractional difference figure (Fig. \ref{fig:frac_diff_pc}) in Appendix \ref{app:peaks}. Across both statistics, we observe several overarching trends.

First, the magnitude of the baryonic bias is strongly scale-dependent. The finest scale ($j=1$, left panels) exhibits the most significant deviations, consistent with the physical picture that baryonic feedback operates primarily on small scales. As we move to larger filter scales ($j=2,3$), the fractional difference diminishes significantly. By the fourth scale (not shown), the bias becomes completely negligible, allowing us to treat these larger scales as ``safe'' for purely gravitational inference.

Furthermore, both HOS allow us to localize the baryonic bias in signal space, revealing a distinct SNR dependence.
In the noise-dominated bulk regime ($|\nu| \lesssim 2.5$), the fractional difference for both statistics is effectively driven toward zero. In these intermediate environments, the underlying differences induced by baryonic feedback are overwhelmed by the variance of the invariant galaxy shape noise \citep{martinet_probing_2021}. Moving to the signal-dominated positive tail, we observe a significant fractional suppression, where the baryonified maps exhibit a clear deficit of high-SNR pixels compared to the DMO baseline. At the extreme positive tail ($\nu \gtrsim 6$), these structures are intrinsically scarce, and the statistical uncertainties grow accordingly. Similarly, in the negative tail, we observe a minor baryonic suppression, which is however also completely washed out by the statistical variance. 
Because the negative tail of the distribution truncates abruptly in deep under-dense regions, the absolute signal quickly becomes identically zero at the lowest SNRs where the starlet $\ell_1$-norm is empty.

To disentangle these noise-induced artifacts from the true underlying signature, we present the same HOS measurements extracted from strictly noiseless maps in Appendix \ref{app:noiseless_stats} (Fig. \ref{fig:frac_diff_pc_noiseless}, \ref{fig:frac_diff_l1_noiseless}).

In principle, one could mitigate the baryonic bias in HOS by excluding only the specific SNR bins most affected by baryons (e.g., discarding $\nu > 2.5$) while retaining the rest of the distribution \citep{huang2019modelling, giblin2018kids}. However, determining the optimal balance between removing contaminated bins and retaining the clean cosmological information potentially entangled within those same ranges is non-trivial. Therefore, we adopt a simple and robust ``scale cut'' strategy: we eschew fine-tuned cuts in SNR space and instead exclude entire wavelet scales (e.g., discarding the $j=1$ maps entirely) if their overall bias contribution exceeds our tolerance threshold.
We leave the complex optimization of fine-grained SNR-space cuts for future investigation.

\subsection{Mitigation technique: the BNT transform} \label{sec:bnt}

To mitigate the impact of baryonic feedback, we investigate the efficacy of the Bernardeau-Nishimichi-Taruya (BNT) transform \citep{bernardeau_cosmic_2014}. The BNT transform is a linear nulling scheme designed to reorganize tomographic data, decomposing the projected weak lensing signal into components that are localized in redshift.

In standard weak lensing tomography, the convergence signal in the $i$-th redshift bin is a line-of-sight projection of the matter density contrast $\delta$, weighted by a broad lensing efficiency kernel $q_i(\chi)$: 
\begin{equation} 
    \kappa_i(\hat{\mathbf{n}}) = \int_0^{\chi_{\mathrm{H}}} \mathrm{d}\chi \, q_i(\chi) \, \delta(\chi \hat{\mathbf{n}}). 
\end{equation} 
As illustrated in Figure \ref{fig:kernels} (left panel) in Appendix \ref{app:bnt_viz}, these broad kernels cause significant mixing of physical scales. A fixed angular multipole $\ell$ receives contributions from both linear large-scale structures at high redshifts and non-linear small-scale structures at low redshifts. 
This mixing is particularly problematic for baryonic mitigation. Baryonic feedback is a physical process restricted to small, non-linear physical scales (high $k$) at relatively low redshifts. In the standard basis, cutting high-$\ell$ modes to remove this low-$z$ contamination effectively discards the clean, linear information coming from high-$z$ lenses at those same angular scales.

The BNT transform addresses this by constructing a new set of transformed convergence fields, $\hat{\kappa}_i$, via a linear combination of the original tomographic bins: 
\begin{equation} 
    \hat{\kappa}_i (\hat{\mathbf{n}}) = \sum_{j=1}^{N_{\mathrm{tomo}}} M_{ij} \, \kappa_j(\hat{\mathbf{n}}). 
\end{equation} 
The transformation matrix $M$ is constructed such that the effective lensing kernels of the transformed fields,
\begin{equation} 
    \hat{q}_i(\chi) = \sum_{j=1}^{N_{\mathrm{tomo}}} M_{ij} \, q_j(\chi), 
\end{equation} 
vanish for lenses located at redshifts lower than the median redshift of bin $i-1$. This nulling property, visualized in Figure \ref{fig:bnt_kernels} (right panel), effectively ``de-projects'' the signal, isolating the contribution of specific source planes and reducing the sensitivity to foreground structures.

Following the prescription in \citet{bernardeau_cosmic_2014}, the coefficients $M_{ij}$ are derived from the geometry of the survey and the source galaxy distributions $n(z)$. The matrix is triangular (typically $M_{ij}=0$ for $j>i$) and normalized such that the diagonal elements are unity. For the discrete case, the coefficients are determined by the requirement to cancel the lensing efficiency factors  $(\chi_s - \chi_l) / \chi_s$ for foreground lenses. The first $z$-bin typically remains untransformed (or simply rescaled). For subsequent $z$-bins $i>1$, the transformation subtracts weighted combinations of lower-redshift bins to cancel out the lensing signal accumulated up to that distance.

While BNT can be applied directly to power spectra or shear catalogs, we apply the transformation at the map level (pixelized HEALPix convergence maps). This enables us to measure non-Gaussian statistics (starlet peak counts and starlet $\ell_1$-norm) directly on the BNT-basis maps.
By suppressing the sensitivity to low-redshift structure, where the angular baryonic signal is strongest, the BNT transform theoretically allows for the inclusion of smaller angular scales (higher $\ell_{\rm max}$) without incurring baryonic bias. 

To visualize the impact of this transformation on the field properties, we provide a comparison of the standard and BNT-transformed mass maps in Appendix \ref{app:bnt_viz}, Figure \ref{fig:noiseless_tomo_maps_stacked_comparison}. 

\begin{figure*}[t]
    \centering
    \includegraphics[width=\textwidth]{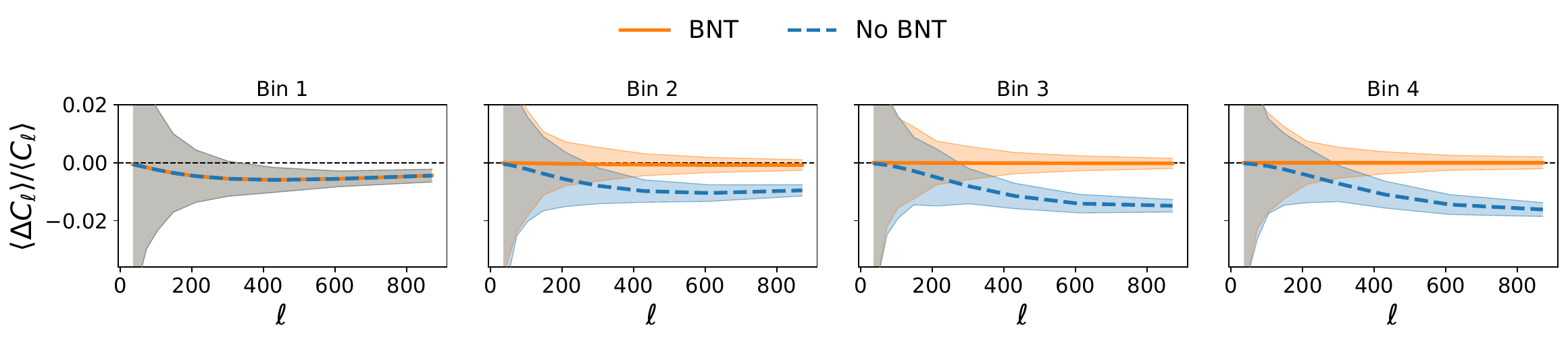}
    \caption{
    Fractional difference in the PS between baryonified and DMO full-sky maps across the four $z$-bins (left to right), comparing the standard and BNT-transformed fields, with shape-noise realizations matched as in Fig.~\ref{fig:frac_diff_ps} and additionally shared between the two bases. Lines and shaded bands show the mean fractional difference and the $1\sigma$ statistical uncertainty of the measured spectrum for the standard basis (blue) and the BNT basis (orange). In bin 1 the two bases coincide by construction.
    } \label{fig:frac_diff_ps_bnt}
\end{figure*}

\begin{figure*}[t]
    \centering
    \includegraphics[width=\textwidth]{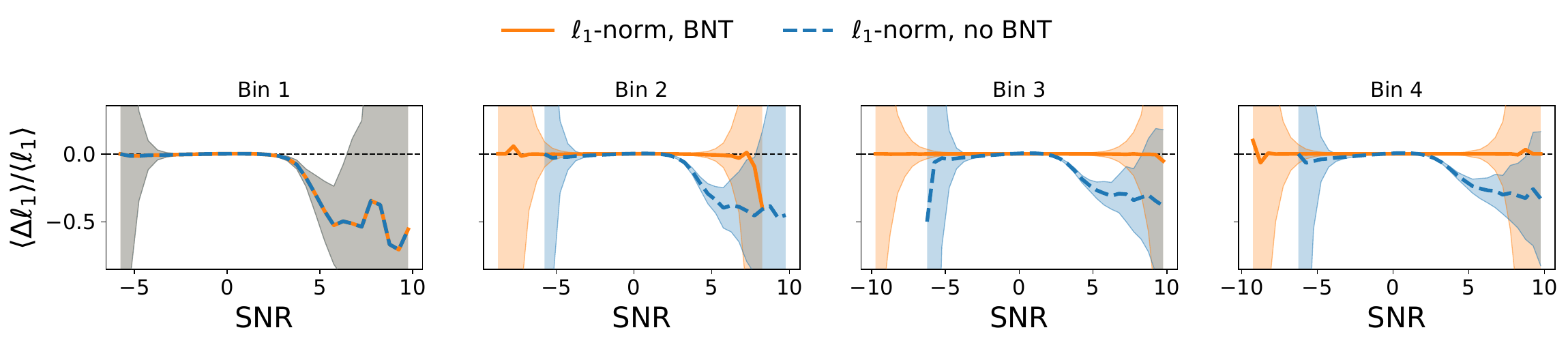}
    \caption{
    Fractional difference in the starlet $\ell_1$-norm (first wavelet scale) between baryonified and DMO full-sky maps as a function of SNR across the four $z$-bins, with shape-noise realizations matched as in Fig.~\ref{fig:frac_diff_l1} and additionally shared between the two bases. Standard tomographic measurements are shown in blue and measurements on the BNT-transformed maps in orange, with shaded bands giving the $1\sigma$ statistical uncertainty of the measured statistic. Bins in which the DMO statistic vanishes are masked. In bin 1 the two bases coincide by construction.
    }
    \label{fig:frac_diff_l1_bnt}
\end{figure*}

Figures \ref{fig:frac_diff_ps_bnt} and \ref{fig:frac_diff_l1_bnt} quantify the performance of the BNT transform for the PS and the starlet $\ell_1$-norm, respectively.
Because the starlet peak counts exhibit a qualitatively similar response as the starlet $\ell_1$-norm, we present their corresponding figure (Fig. \ref{fig:frac_diff_pc_bnt}) in Appendix \ref{app:peaks}.
Across all three statistics, the transform is highly effective at removing the mean baryonic bias. 
For the PS, for $z$-bins 2, 3, and 4, the fractional difference (orange lines) is flattened almost perfectly to zero across the entire multipole range. For the non-Gaussian statistics, while the mean fractional difference does not vanish completely at certain SNR values, the residual shifts remain well within the 1$\sigma$ statistical uncertainty, rendering them statistically insignificant.

The transform comes at a cost in signal-to-noise. Nulling is achieved by subtracting the tomographic maps from one another. The signal they share cancels, which is the intent, but their shape noise is independent between bins, so it does not cancel and the two noise variances add. Each transformed map at higher redshift therefore carries less signal and more noise than the tomographic map it is built from, and the shape noise, originally independent, becomes correlated between the transformed maps. Appendix \ref{app:bnt_viz} (Fig. \ref{fig:noisy_tomo_maps_stacked_comparison}) shows this directly on the maps. It does not show in Figs. \ref{fig:frac_diff_ps_bnt} and \ref{fig:frac_diff_l1_bnt}, where each band is a relative uncertainty, the scatter of a measurement divided by the mean of the same measurement in the same basis, and is therefore insensitive to an overall change in the amplitude of the statistic. The consequences for the inferred parameters are quantified in Sect. \ref{sec:bnt_results}.

\section{Inference methodology} \label{sec:inference}

Standard cosmological parameter inference relies on the evaluation of the likelihood function, $\mathcal{L}(\boldsymbol{x}|\boldsymbol{\theta}) \equiv p(\boldsymbol{x}|\boldsymbol{\theta})$, which quantifies the probability of observing the data vector $\boldsymbol{x}$ given a physical model parameterised by $\boldsymbol{\theta}$. In the regime of Gaussian random fields, this likelihood is analytically tractable and fully characterized by the mean and covariance of the data. However, for the analysis of non-Gaussian descriptors of the cosmic web, such as starlet peak counts and the starlet $\ell_1$-norm, the assumption of Gaussianity may break down.

For these HOS, the likelihood function is unknown and analytically intractable due to the non-linear evolution of the density field and the complex interplay of observational systematics (e.g., survey masks and shape noise).
Furthermore, even for independent probes, the high dimensionality of multiscale data vectors makes analytical covariance matrices difficult to estimate robustly from a finite set of simulations \citep{hartlap_why_2007, taylor2013putting}.

To overcome these limitations, we adopt a simulation-based inference (SBI) approach, also referred to as likelihood-free inference \citep{cranmer_frontier_2020}. SBI bypasses the need for an explicit analytic likelihood. Instead, it treats the entire forward modeling pipeline (in our case, the N-body simulation pipeline plus the summary statistic measurement) as a stochastic simulator that generates samples $\boldsymbol{x} \sim p(\boldsymbol{x}|\boldsymbol{\theta})$ given input parameters $\theta$. Using the \texttt{cosmoGRID} suite,
we can sample from the joint distribution of cosmological parameters and observables, $p(\theta,x)$, using these samples to directly estimate the posterior 
while implicitly capturing the full non-Gaussian information and the complex joint covariance of our probes.

\subsection{Simulation-based inference (SBI) with neural posterior estimation} \label{sec:sbi_npe}

To infer cosmological parameters, we employ Neural Posterior Estimation (NPE; \citealt{papamakarios_fast_2016}). 
NPE requires a set of training pairs $(\boldsymbol{\theta}_i, \boldsymbol{x}_i)$ generated by first
drawing parameters from a prior distribution, $\boldsymbol{\theta}_i \sim p(\boldsymbol{\theta})$, and then producing
the corresponding simulated data vector $\boldsymbol{x}_i \sim p(\boldsymbol{x}|\boldsymbol{\theta}_i)$ through
the forward model. A neural network is then trained to directly approximate the posterior distribution
$p(\boldsymbol{\theta}|\boldsymbol{x})$ from these simulated pairs.

A major advantage of NPE is that it is amortized: once the network is trained, the posterior for any newly observed data vector can be evaluated almost instantaneously, bypassing the need for computationally expensive MCMC sampling steps. 
This computational efficiency is especially important for the systematic study presented in this work, as we are required to perform inference hundreds of times to rigorously test different combinations of survey areas, summary statistics, and scale cuts.

We estimate the posterior using a conditional density estimator $q_\phi(\boldsymbol{\theta}|\boldsymbol{x})$, parameterized by a neural network with weights $\phi$. The network is trained to minimize the forward Kullback-Leibler (KL) divergence between the true posterior and the estimate. 
This objective is equivalent to maximizing the log-probability of the true parameters under the network's predicted distribution, averaged over the joint distribution of the training simulations:
\begin{equation} \label{eq:npe_loss}
    \mathcal{L}(\phi) = - \mathbb{E}_{p(\boldsymbol{\theta}, \boldsymbol{x})} \left[ \log q_{\phi}(\boldsymbol{\theta} | \boldsymbol{x}) \right].
\end{equation}
In practice, this expectation is approximated by averaging over
mini-batches of simulated realizations from the training set. We note
that the prior $p(\boldsymbol{\theta})$ enters only implicitly through
the distribution of training parameters: the learned
$q_\phi$ approximates the posterior with respect to whatever prior was
used to generate the training set. In our case, the prior is defined by
the \texttt{cosmoGRID} sampling design, which draws cosmologies from an
approximate uniform distribution over the parameter ranges listed in
Table~\ref{tab:cosmogrid_priors} using a Sobol quasi-random sequence
\citep[see][]{kacprzak_cosmogridv1_2023}.

\subsection{Density estimation with masked autoregressive flows} \label{sec:maf}

To model the complex conditional distribution $q_\phi(\boldsymbol{\theta} | \boldsymbol{x})$, we use normalizing flows \citep{rezende_variational_2015}, specifically the masked autoregressive flow (MAF) architecture \citep{papamakarios_masked_2017}. 

A MAF constructs the target posterior density by transforming a simple base distribution $p(\boldsymbol{u})$ (typically a standard multivariate Gaussian, $u \sim \mathcal{N}(0, \mathbb{I})$) through a series of invertible and differentiable transformations $f_\phi$. 
The target parameters are modeled as $\boldsymbol{\theta}=f_\phi(\boldsymbol{u};\boldsymbol{x})$, and the density is evaluated using the change of variables formula: 
\begin{equation} 
    q_{\phi}(\boldsymbol{\theta} | \boldsymbol{x}) = p(\boldsymbol{u}) \left| \det \frac{\partial f_{\phi}^{-1}(\boldsymbol{\theta}; \boldsymbol{x})}{\partial \boldsymbol{\theta}} \right|. 
\end{equation} 
The MAF architecture is particularly well-suited for this task as it employs autoregressive layers (masked autoencoders for distribution estimation; MADE) to define the transformation $f_\phi$. This structure ensures that the Jacobian determinant is triangular and computationally efficient to evaluate, enabling scaling to high-dimensional parameter spaces.

\subsection{Implementation and training details} \label{sec:implementation}

We implement our inference pipeline using \texttt{jaxili}\footnote{\url{https://github.com/sachaguer/jaxili}}, a Python package for SBI built upon the JAX high-performance numerical computing library \citep{jax2018github}. JAX provides automatic differentiation and Just-In-Time (JIT) compilation, allowing for highly efficient training of the flow networks on GPUs.

\subsubsection{Data pre-processing and augmentation}

The input data vector $\boldsymbol{x}$ is constructed by concatenating the raw data vectors for our summary statistics, as described in Sect. \ref{sec:stats}. Given the dimensionality of our summary statistics, we feed the measured data vectors directly into the network without an intermediate score compression step, relying on the neural network to learn the optimal feature extraction. To facilitate network convergence, we standardize both the input summary statistics and the target cosmological parameters to have zero mean and unit variance based on the training set statistics. The data is randomly split into 70\% training, 20\% validation, and 10\% test sets.

The networks are trained on the 2,500 cosmologies from the \texttt{cosmoGRID} suite. To capture the cosmic variance inherent in the estimators, for each cosmology we have access to 7 realizations (via shell permutations). To robustly model the impact of galaxy shape noise, we augment the training set by generating multiple independent shape noise realizations for each cosmological simulation. This combination results in a total of $\sim$52,500 training pairs, ensuring the network accurately learns the scatter in the observables due to both cosmic variance and noise.

\subsubsection*{Network architecture and training}
We use a MAF consisting of 5 autoregressive transformations. 
Each transformation is parameterized by a MADE network with 2 hidden layers of 50 units each, using ReLU activation functions. The input ordering is reversed between successive transformations.

Training is performed using the Adam optimizer with a peak learning rate of $10^{-4}$ and a warmup cosine decay schedule. We employ early stopping based on the validation loss with a patience of 20 epochs and a minimum improvement threshold of $10^{-3}$, with a maximum of 1,000 epochs. Gradients are clipped by global norm at $5.0$. The training batch size is 40.

\subsubsection*{Validation}

To assess the reliability of our inferred posteriors, we perform simulation-based coverage tests using the \texttt{TARP} package \citep{lemos_sampling-based_2023}\footnote{\url{https://github.com/Ciela-Institute/tarp}}. 
This diagnostic evaluates whether the posterior credible regions possess the correct frequentist coverage. By running the trained estimator on an independent ensemble of simulated test datasets, we compute the empirical coverage probability (ECP) as a function of the nominal credibility level $\alpha$. We require that $\mathrm{ECP}(\alpha) \approx \alpha$ (e.g., the $68\%$ credible region should contain the true parameters in $68\%$ of the test simulations). We rigorously applied this validation step for every distinct posterior estimator trained in this study to ensure that all reported contours are statistically well-calibrated.

\subsection{Impact of survey area: experimental setup} \label{sec:survey_area_setup}

The detectability of baryonic biases depends critically on the statistical constraining power of the survey, which scales with the observed sky fraction ($f_{\rm sky}$).
As the survey area increases, the statistical uncertainties (contours) shrink 
roughly as $\propto 1/ \sqrt{f_{\rm sky}}$ \citep{dodelson2020modern}, which we verify on our own posteriors.
Consequently, a systematic shift that is statistically negligible for a Stage III dataset may manifest as a catastrophic, multi-sigma tension for a high-precision Stage IV experiment.

To quantify this scaling and derive survey-specific recommendations for scale cuts, we perform our inference analysis across a range of survey areas.
We focus on four primary configurations: a baseline area (2,000 deg$^2$) representative of early data releases; a Stage III-like footprint (5,000 deg$^2$) approximating surveys like DES or KiDS; a Stage IV-like footprint (14,000 deg$^2$) anticipating missions like \textit{Euclid} and LSST; and the full sky ($\approx$41,253 deg$^2$) to establish the theoretical limit ($f_{\rm sky}=1$). To densely map the evolution of the bias, we also evaluate three intermediate areas: 10,000, 28,000, and 35,000 deg$^2$.

To maintain generality and ensure our results depend primarily on the survey area rather than the specific geometry of a given instrument's footprint, we employ a generic binary masking scheme. For each targeted sky fraction, we apply a binary mask to the full-sky HEALPix maps, retaining a contiguous polar cap region corresponding to the desired solid angle. While real survey footprints are complex and disjoint, this naive masking approach preserves the fundamental statistical properties of the field (sample variance) and allows us to isolate the dependency of the baryonic bias solely on the available information content.

Under a forward model, a summary statistic is meaningful whatever the mask does to it, provided the same operation is applied to the data and to the simulations. We nevertheless decouple the power spectrum, computing binned bandpowers with \texttt{NaMaster}, which inverts the mode-coupling matrix induced by the mask geometry. At fixed binning this is an invertible linear transformation of the pseudo-$C_\ell$, so it neither adds nor removes information. What it provides is a multipole label that means the same thing at every footprint, so that the scale cuts of Sect. \ref{sec:scale_cuts} and the area scaling of Sect. \ref{sec:bias_vs_area} can be compared across survey configurations. We also subtract the mean convergence inside the footprint from each map. That mean is not observable from shear, so the simulations carry an absolute zero point that no real measurement has, and we remove it from both. The mask spreads this mode over a range of multipoles, so subtracting it keeps the low-$\ell$ end of our range usable. For the HOS we apply the mask before the starlet transform and measure the coefficients within the footprint. The detail coefficients are compensated and so insensitive to a constant offset, but masking leaves a step at the survey boundary to which the filter does respond, and subtracting the mean removes the part of that step that varies with cosmology.

For each survey configuration and each summary statistic (PS, starlet peak counts, starlet $\ell_1$-norm), we repeat the full SBI pipeline. This systematic variation allows us to determine the ``tipping point'' where unmodeled baryonic feedback induces a statistically significant bias ($>0.3\sigma$) in the posterior contours, thereby establishing area-dependent scale cut prescriptions.

\section{Inference results} \label{sec:results}

\subsection{The scaling of baryonic bias with survey area} \label{sec:bias_vs_area}

We begin by quantifying the magnitude of the bias introduced by unmodeled baryonic feedback when using the full information content available at the resolution limit of our maps. 
For the PS, we include multipoles $37 \leq \ell \leq 1024$. The upper bound corresponds to the standard $2N_{\rm side}$ sampling limit required to minimize aliasing and pixel window effects, and the lower bound matches the multipole coverage of the wavelet bands entering the HOS data vector (Appendix~\ref{app:starlet_ell}). For the HOS (starlet peak counts and the $\ell_1$-norm), we include all four wavelet scales, down to the finest band available at $N_{\rm side}=512$.

Using a density estimator trained exclusively on DMO simulations, we derive posteriors for two ``observed'' data vectors: an unbiased reference (DMO fiducial) and a biased target (baryonified fiducial).
We quantify the baryonic bias as the statistical tension between these posteriors using the difference-of-means statistic, $Q_{\rm DM}$, via the \texttt{tensiometer} package \citep{raveri_quantifying_2020, raveri2021non}. This metric provides a parameter-reparametrization-invariant measure of agreement between two posteriors, $p_1$(unbiased baseline) and $p_2$ (baryon-biased), while properly accounting for the non-Gaussian nature of the parameter space and the information content of the prior.

The tension is quantified by the quadratic form: 
\begin{equation} 
    Q_{\rm DM} \equiv (\boldsymbol{\theta}_1-\boldsymbol{\theta}_2) \, \mathcal{C}_{\rm shift}^{-1} \, (\boldsymbol{\theta}_1-\boldsymbol{\theta}_2)^T, 
\end{equation}
where $\boldsymbol{\theta}_{1,2}$ are the posterior means. Because both experiments share the same underlying SBI prior, standard tension metrics underestimate the significance of the shift. We therefore use the exact shift covariance definition that corrects for shared prior volume: 
\begin{equation} 
    \mathcal{C}_{\rm shift} = \mathcal{C}_{p_1} + \mathcal{C}_{p_2} - \mathcal{C}_{p_1}\mathcal{C}_{\Pi}^{-1}\mathcal{C}_{p_2} - \mathcal{C}_{p_2}\mathcal{C}_{\Pi}^{-1}\mathcal{C}_{p_1}, 
\end{equation} 
where $\mathcal{C}_{p_1, p_2}$ are the posterior covariances and $C_\Pi$ is the prior covariance.

We restrict the calculation to the subspace of directions actually constrained by the weak lensing data ($\Omega_m, \sigma_8, w_0$), excluding unconstrained projection parameters ($H_0, n_s, \Omega_b$). The resulting $Q_{\rm DM}$ value is converted into a Gaussian-equivalent statistical significance ($\sigma$), which we report as the ``baryonic bias''.

\begin{figure}[t]
    \centering
    \includegraphics[width=0.48\textwidth]{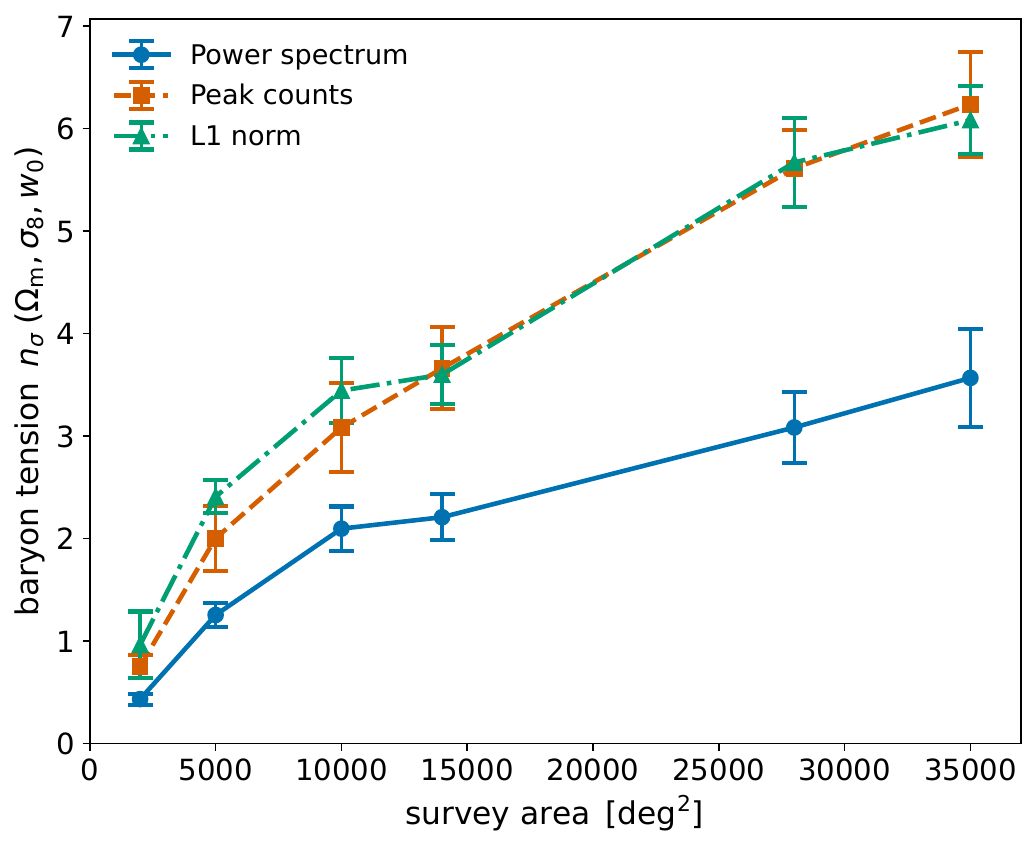}
    \caption{
    Scaling of the baryonic bias ($n_\sigma$) as a function of survey area for the PS (blue circles), starlet peak counts (orange squares), and starlet $\ell_1$-norm (green triangles), evaluated at full map resolution ($\ell_{\rm max} = 1024$ and all four wavelet scales). Data points and error bars correspond to the mean and standard deviation across independent NPE training runs.
    }
    \label{fig:nsigma_vs_area}
\end{figure}

Figure \ref{fig:nsigma_vs_area} presents the evolution of this baryonic bias for the PS, starlet peak counts, and starlet $\ell_1$-norm as a function of survey area. To ensure the robustness of these results against training stochasticity, each data point represents the average over five independent NPE training runs, with error bars indicating the standard deviation across these runs.

\begin{figure}[t]
    \centering
    \includegraphics[width=0.48\textwidth]{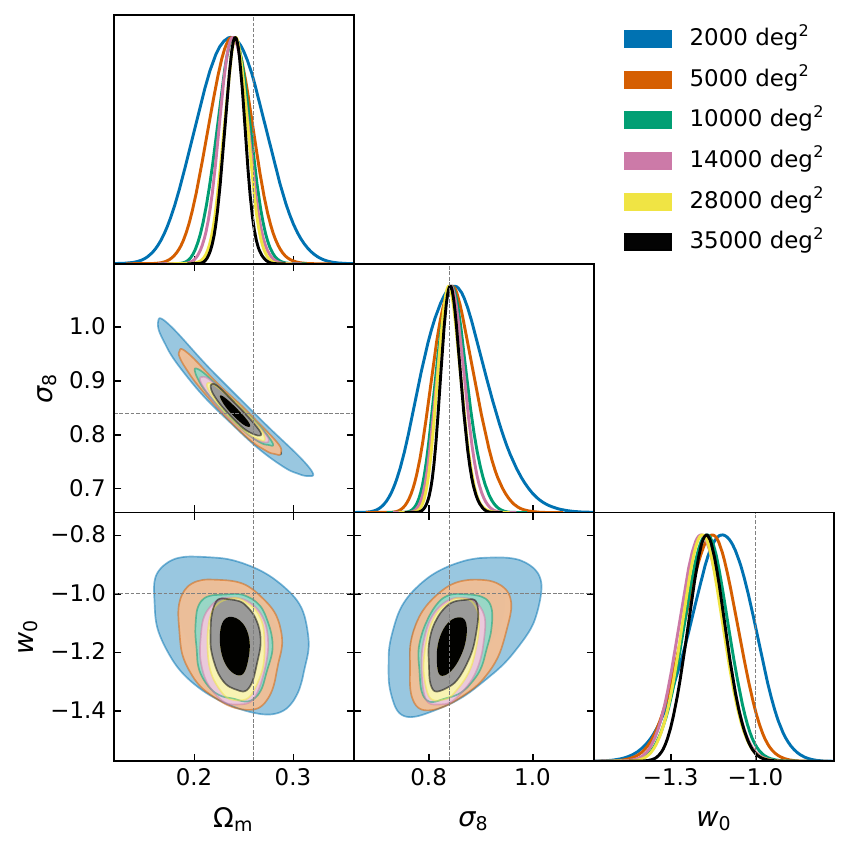}
    \caption{
        Posterior contours in the ($\Omega_m, \sigma_8, w_0$) plane derived from the PS at full resolution ($\ell_{\rm max}=1024$) for the baryonified mock data. The contours are shown for six survey areas ranging from $2{,}000$ to $35{,}000$ deg$^2$, with dashed lines marking the true parameter values.
    }
    \label{fig:PS_contours_vs_area}
\end{figure}

\begin{figure}[t]
    \centering
    \includegraphics[width=0.48\textwidth]{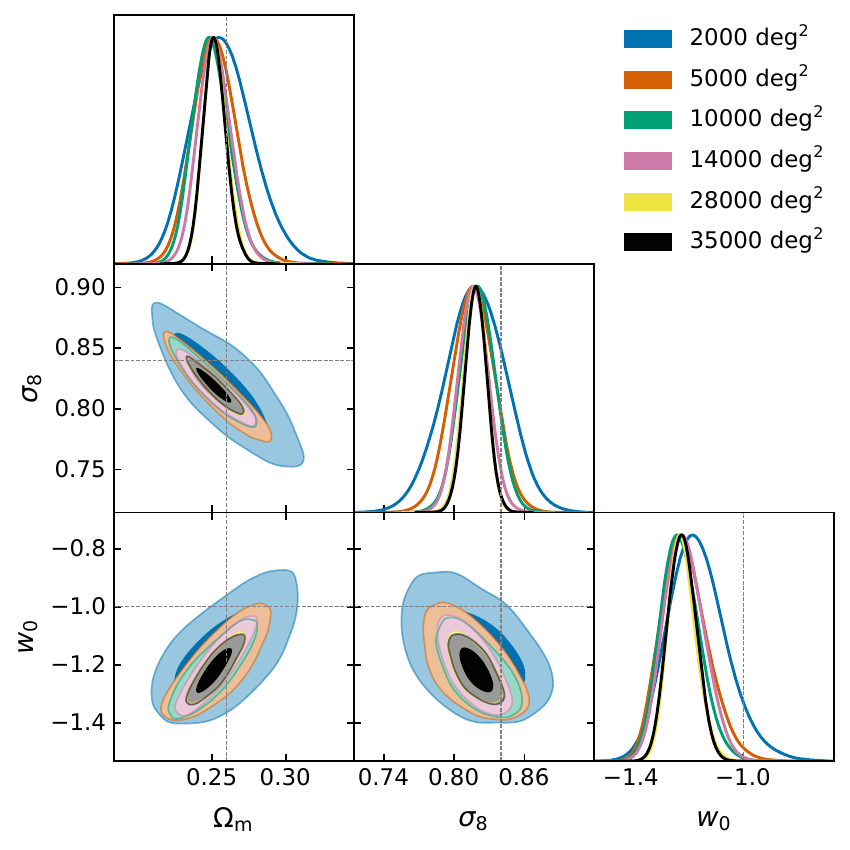}
    \caption{
    Same as Fig. \ref{fig:PS_contours_vs_area}, but derived from the starlet $\ell_1$-norm at full map resolution (including the finest wavelet scale, $j=1$).
    } 
    \label{fig:l1_contours_vs_area}
\end{figure}

At full map resolution, the bias exceeds our strict tolerance threshold ($0.3\sigma$) even for the smallest footprint considered. For the baseline 2,000 deg$^2$ area, the PS exhibits a bias of $\sim 0.4\sigma$, while the starlet peak counts and the starlet $\ell_1$-norm already show stronger deviations of $\sim 0.8\sigma$ and $\sim 1.0\sigma$, respectively.
As the survey area expands, the statistical precision improves, exposing the systematic shift with increasing significance. For a Stage IV-like area (14,000 deg$^2$), the tension in the PS rises to $\sim 2.2\sigma$ and both wavelet statistics reach $\sim 3.6\sigma$, indicating statistically significant deviations between the inferred and true parameters.

For survey areas approaching the full sky (>30,000 deg$^2$), the tension becomes highly significant across all probes. The bias for the PS reaches $\sim 3.5\sigma$, while both wavelet statistics exhibit parameter shifts exceeding $6\sigma$. This confirms that for future wide-field surveys, unmodeled baryonic feedback constitutes a major systematic uncertainty, necessitating explicit mitigation strategies to recover accurate cosmological constraints. 

Comparing the different statistics at this fixed resolution limit, the two wavelet statistics exhibit very similar levels of bias, consistently higher than that of the PS. The origin of this difference is visible in the posterior displacements (Figs. \ref{fig:PS_contours_vs_area} and \ref{fig:l1_contours_vs_area}). The two wavelet statistics respond to the unmodelled feedback with nearly identical displacements in parameter space, shifting $\sigma_8$ and $w_0$ low with only a small displacement in $\Omega_m$. The PS instead slides along its $\Omega_m$--$\sigma_8$ degeneracy toward lower $\Omega_m$, leaving $\sigma_8$ nearly unbiased, and absorbs the remainder of the shift in $w_0$. Combined with the tighter contours of the wavelet statistics, these displacements translate into the larger statistical tensions of Fig. \ref{fig:nsigma_vs_area}.

\subsection{Determining robust scale cuts} \label{sec:scale_cuts}

Having quantified the impact of unmodeled baryons on the full data vector, we now turn to the practical question: \textit{What range of scales can be safely used for cosmological inference without explicit baryonic modeling?}

We define a ``robust'' scale cut as the threshold (either the maximum multipole $\ell_{\rm max}$ for the PS or the minimum smoothing scale $\theta_{\rm min}$ for HOS) at which the baryonic bias becomes statistically negligible. We adopt a strict convergence criterion, requiring the tension between the baryonified and DMO posteriors (calculated as explained in Sect. \ref{sec:bias_vs_area}) to drop below $0.3\sigma$ (i.e., bias $<0.3$) as measured by the $Q_{\rm DM}$ metric in the ($\Omega_m, \sigma_8, w_0$) subspace. This strict threshold ensures that the systematic error is negligible compared to the statistical precision of the survey.

We perform an iterative optimization search: 
\begin{itemize} 
    \item PS: We progressively reduce $\ell_{\rm max}$ from 1020 downwards in steps of $\Delta \ell=40$.
    \item HOS: We progressively exclude the finest wavelet scales. The ``scale cut'' is defined by the finest wavelet scale retained in the analysis.
\end{itemize}
For the PS, we adopt as the scale cut the largest grid value of $\ell_{\rm max}$ for which the bias remains below, or statistically consistent with, the threshold.

\begin{figure*}[t]
    \centering
    \includegraphics[width=\textwidth]{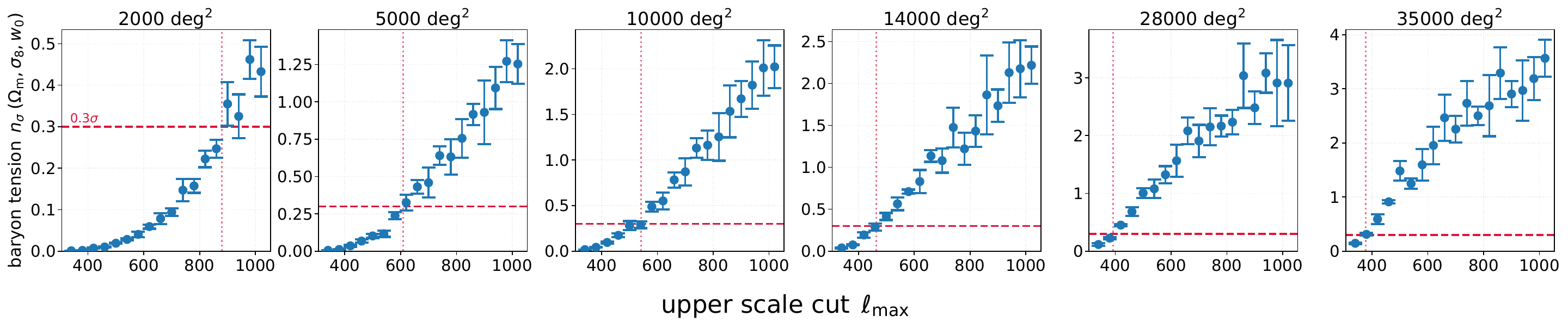}
    \caption{
    Baryonic bias ($n_\sigma$) of the PS as a function of the maximum multipole $\ell_{\rm max}$ included in the inference. Panels represent different survey footprints, increasing in area from left to right. Data points and error bars show the mean and standard deviation across independent NPE training runs. The horizontal red line marks the $0.3\sigma$ threshold used to define the safe $\ell_{\rm max}$ cut, and the vertical dotted line marks the interpolated crossing of this threshold in each panel.
    }
    \label{fig:nsigma_vs_cut}
\end{figure*}

Figure \ref{fig:nsigma_vs_cut} illustrates this optimization for the PS. It plots the baryonic bias ($n_\sigma$) as a function of the included maximum multipole. As more small-scale information is added (moving right on the x-axis), the bias rises monotonically, crossing the red tolerance threshold (0.3$\sigma$) at different points depending on the survey area.

\begin{table}[t]
\centering
\begin{tabular}{c c c}
\hline \hline
\textbf{Survey area [deg$^2$]} & \textbf{PS ($\ell_{\rm max}$)} & \textbf{HOS (scales)} \\
\hline
2,000 & 860 & \multirow{7}{*}{$j \geq 2$} \\
5,000 & 580 & \\
10,000 & 540 & \\
14,000 & 460 & \\
28,000 & 380 & \\
35,000 & 380 & \\
Full sky & 340 & \\
\hline
\end{tabular}
\caption{
Required scale cuts to maintain baryonic bias below the $0.3\sigma$ threshold as a function of survey area. For the PS, values give the maximum multipole ($\ell_{\rm max}$) retained in the data vector. For the map-based HOS (starlet peak counts and $\ell_1$-norm), the required cut is the removal of the finest wavelet scale ($j=1$) for every survey configuration.
}
\label{tab:scale_cuts}
\end{table}

The resulting safety thresholds for each statistic and survey configuration are summarized in Table \ref{tab:scale_cuts}.
For a full-sky analysis, the PS requires aggressive scale cuts. To satisfy the $0.3\sigma$ criterion, we must discard all information above $\ell \approx 340$. This removes the majority of the modes typically used in weak lensing to constrain cosmology, severely limiting the statistical power of the probe.
In contrast, we find that for starlet peak counts and the starlet $\ell_1$-norm, removing just the single finest wavelet scale ($j=1$) is sufficient to control the bias for all survey areas. The remaining scales, which dominate multipoles $\ell \lesssim 336$ (Appendix~\ref{app:starlet_ell}), appear remarkably robust to baryonic effects, with the residual bias reaching the $0.3\sigma$ threshold within its uncertainty only at full sky.
    
This result highlights a fundamental advantage of wavelet-based statistics. The starlet transform provides physical scale separation: the baryonic signal, which is physically concentrated on small scales, is effectively isolated into the finest wavelet coefficient map. By discarding this specific map, we precisely remove the contamination.
   
The cuts in Table \ref{tab:scale_cuts} are derived for the feedback prescription implemented in \texttt{cosmoGRID}. A stronger suppression could move them to larger scales.

We note that the HOS scale cuts presented here are inherently conservative due to the dyadic nature of the standard starlet decomposition. Because the characteristic sizes of the wavelet bands increase by powers of two, our ability to fine-tune the spatial filter is highly quantized. As summarized in Table \ref{tab:scale_cuts}, mitigating the bias requires removing the first wavelet scale ($j=1$) across all survey configurations. For the full-sky configuration, the residual bias sits at the $0.3\sigma$ threshold within its uncertainty. However, for smaller survey footprints with higher statistical tolerance, dropping the entire $j=1$ band likely discards more uncontaminated, quasi-linear information than strictly necessary. 
In principle, employing a filter bank with intermediate scale steps, such as Feauveau's $\sqrt{2}$-resolution decomposition \citep{starck_murtagh_2006} or the non-dyadic scale sampling adopted by \citet{zurcher_towards_2023}, would allow for highly optimized, area-dependent scale cuts, analogous to the $\ell_{\rm max}$ sliding window used for the PS, thereby retaining even more constraining power for smaller footprints. The starlet peak counts and the starlet $\ell_1$-norm also offer a form of mitigation that is unavailable to the PS. Both are binned in SNR as well as in scale, and the baryonic response is concentrated in the positive tail (Sect.~\ref{sec:hos}), so the contamination can be removed where it sits while the rest of the band stays in the data vector. The cut can therefore be targeted more finely than a whole band, which leaves them room to improve on the constraints reported here. We leave the exploration of non-dyadic filter banks and of cuts in SNR to future work.

\subsection{Information content at large scales} \label{sec:info_content}

While the scale cuts derived in Sect. \ref{sec:scale_cuts} ensure robustness against baryonic feedback, they come at the cost of discarding small-scale data. This raises a critical question: if we are forced to restrict our analysis to large, quasi-linear scales to avoid systematics, do HOS still provide any constraining advantage over the PS, or does the field become sufficiently Gaussian that the PS captures all available information?

To address this, we compare the constraining power of the three statistics using strictly ``baryon-safe'' data. We limit each statistic to the specific scale ranges identified in Table \ref{tab:scale_cuts} as robust (e.g., restricting the PS to $\ell \leq 460$ at 14,000 deg$^2$, while retaining wavelet scales $j\geq 2$ for the HOS).

We quantify the constraining power using a figure of merit (FoM) defined in the ($\Omega_m, \sigma_8, w_0$) parameter subspace. We calculate the FoM as the inverse of the volume of the $1\sigma$ posterior ellipsoid (or equivalently via the determinant of the parameter covariance sub-matrix): 
\begin{equation} 
    \mathrm{FoM} = \frac{1}{\sqrt{\det(\mathbf{C}_{\Omega_m, \sigma_8, w_0})}}.
\end{equation} 

Table \ref{tab:fom_results} lists the FoM for the PS, starlet peak counts, and starlet $\ell_1$-norm across the different survey areas, calculated after applying the respective scale cuts mandated by the $0.3\sigma$ bias criterion.
The values in parentheses denote the relative gain factor over the PS ($ \mathrm{FoM}_{\rm HOS} / \mathrm{FoM}_{\rm PS} $).

We find that the starlet $\ell_1$-norm yields tighter constraints than the PS at every survey area on these baryon-safe scales, with an advantage that grows with survey area and reaches a factor of 1.8 at 14,000 deg$^2$ and 2.6 in the full-sky limit. This demonstrates that cosmologically valuable non-Gaussian information persists well into the quasi-linear regime. The starlet peak counts instead underperform the PS at the smaller areas and reach approximate parity with it at Stage IV-like areas. The quantized wavelet cut removes a larger fraction of the peak-count information than the sliding $\ell_{\rm max}$ cut removes from the PS, and this penalty is most severe where the PS cut is loosest.

The same granularity shapes the trend of the ratios in Table~\ref{tab:fom_results} with survey area. The sliding $\ell_{\rm max}$ lets the PS retain exactly the multipole range that is safe at each area, whereas the wavelet statistics discard the entire $j=1$ band everywhere, even where their residual bias lies well below the $0.3\sigma$ threshold. The HOS entries at the smaller areas are therefore conservative, and only in the full-sky limit, where the residual HOS bias reaches the threshold, are both statistics cut to their actual limits. This is a restriction of the dyadic decomposition, not of the statistics themselves (Sect.~\ref{sec:scale_cuts}).

\begin{figure}[t]
    \centering
    \includegraphics[width=0.48\textwidth]{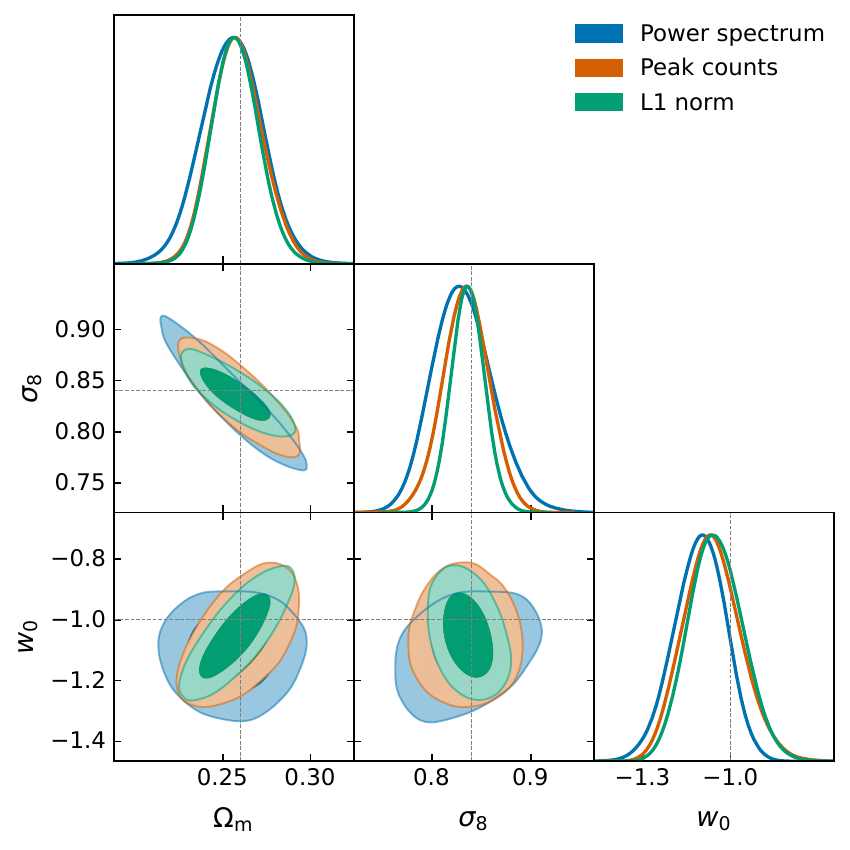}
    \caption{
    Posterior constraints at 14,000 deg$^2$ restricted to baryon-safe scales, for the baryonified mock data. The PS (blue) is truncated at $\ell_{\rm max} = 460$, and the starlet peak counts (orange) and starlet $\ell_1$-norm (green) exclude the finest wavelet scale, retaining scales $j \geq 2$.
    }
    \label{fig:contours_baryon_safe}
    \label{fig:contours_cut_fullsky} % legacy label so old references resolve in the marked-changes diff
\end{figure}

These results imply that the cosmic web imprints detectable, uncorrupted non-Gaussian morphological features on intermediate scales ($j \geq 2$). 
Figure \ref{fig:contours_baryon_safe} shows the three statistics at 14,000 deg$^2$ after the cuts of Table \ref{tab:scale_cuts}. We show this footprint because it is the configuration the Stage IV surveys motivating this work will observe. The $\ell_1$-norm contours remain substantially tighter than the PS contours, while the peak-count contours are comparable in size to them. The orientation of the contours differs as well. In the ($\Omega_m$, $w_0$) and ($\sigma_8$, $w_0$) planes the degeneracy directions of both HOS differ from that of the PS, so the statistics are partly complementary. The peak counts therefore add information despite reaching a figure of merit comparable to that of the PS (Table~\ref{tab:fom_results}). While discarding small scales inevitably reduces the total information available to the survey, the starlet $\ell_1$-norm recovers a significant amount of statistical precision from the remaining large scales.

\begin{table*}[t] 
    \centering 
    \begin{tabular}{l c c c c} 
        \hline \hline
        \textbf{Statistic} & \textbf{2,000 deg$^2$} & \textbf{5,000 deg$^2$} & \textbf{14,000 deg$^2$} & \textbf{Full sky}
        \\ \hline
        PS & $2.3 \pm 0.3$ & $6.4 \pm 0.7$ & $14.5 \pm 0.9$ & $54.9 \pm 2.9$
        \\ Starlet peak counts & $1.1 \pm 0.1$ ({\boldmath$\times 0.46$}) & $4.1 \pm 0.5$ ({\boldmath$\times 0.63$}) & $15.5 \pm 3.5$ ({\boldmath$\times 1.07$}) & $64.0 \pm 7.2$ ({\boldmath$\times 1.17$})
        \\ Starlet $\ell_1$-norm & $2.7 \pm 0.2$ ({\boldmath$\times 1.15$}) & $10.3 \pm 0.4$ ({\boldmath$\times 1.60$}) & $26.1 \pm 8.2$ ({\boldmath$\times 1.80$}) & $143.4 \pm 21.4$ ({\boldmath$\times 2.61$})
        \\ \hline
    \end{tabular}
    \caption{
    Figure of merit (FoM, in units of $10^4$) evaluated in the ($\Omega_m, \sigma_8, w_0$) parameter subspace on baryon-safe scales, applying the scale cuts of Table \ref{tab:scale_cuts}. Uncertainties are the standard deviation across the five independent NPE training runs. Values in parentheses give the FoM ratio of each statistic to the PS.}
    \label{tab:fom_results} 
\end{table*}

\subsection{Performance of the BNT transform} \label{sec:bnt_results}

Finally, we evaluate the efficacy of the BNT transform as an alternative mitigation strategy. As detailed in Sect. \ref{sec:bnt}, the BNT transform reorganizes the tomographic data vector to null the lensing efficiency at low redshifts. Theoretically, this should allow us to isolate the small-scale baryonic contamination to the lowest redshift bin, leaving the high-redshift bins ``clean'' even at small angular scales.

\begin{figure*}[t]
     \centering
     % --- First Subfigure ---
     \begin{subfigure}[b]{0.48\textwidth}
         \centering
         \includegraphics[width=\textwidth]{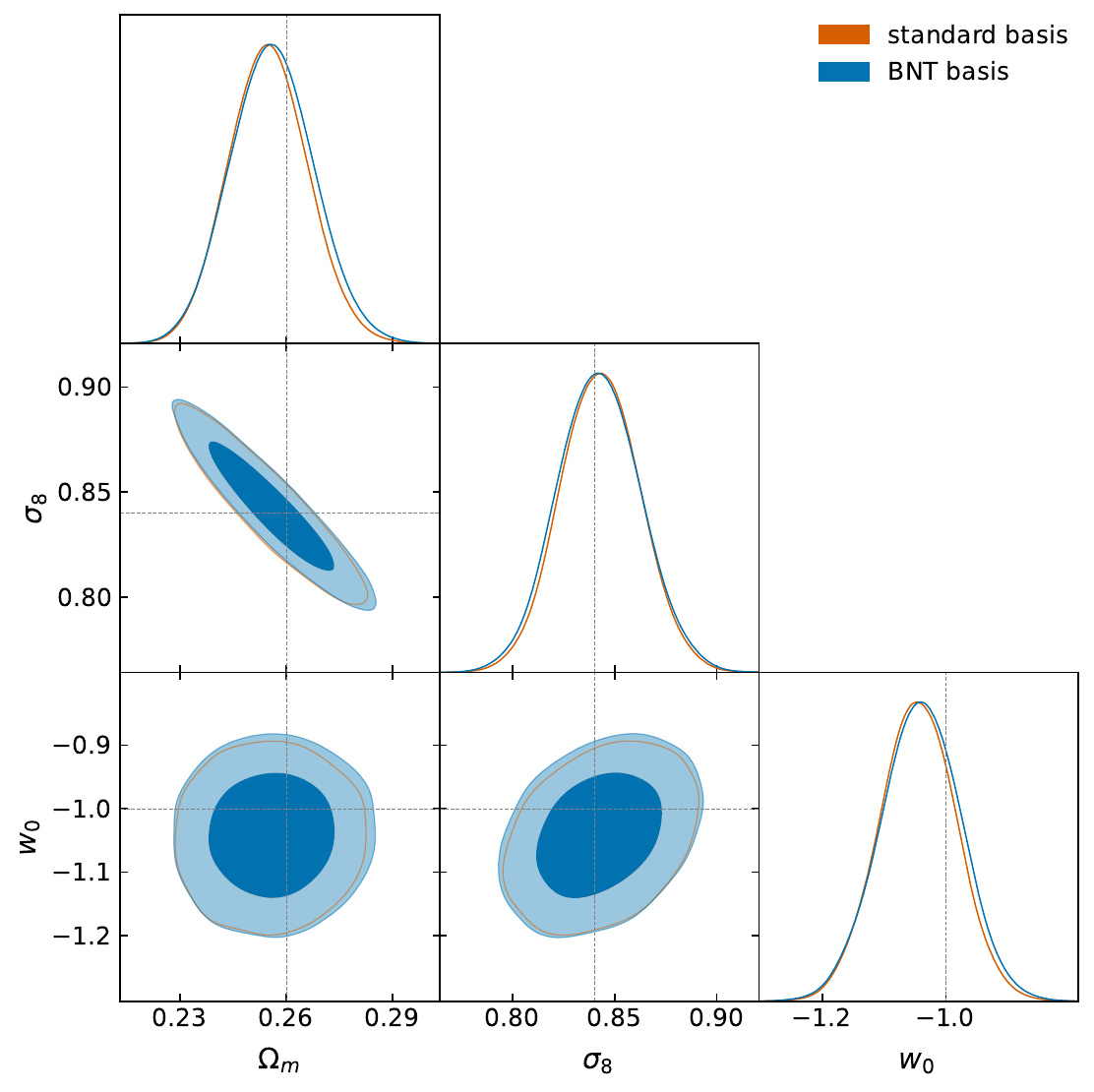}
         \caption{Full map resolution ($\ell_{\rm max}=1024$), no scale cuts.}
         \label{fig:bnt_1024}
     \end{subfigure}
     \hfill
     % --- Second Subfigure ---
     \begin{subfigure}[b]{0.48\textwidth}
         \centering
         \includegraphics[width=\textwidth]{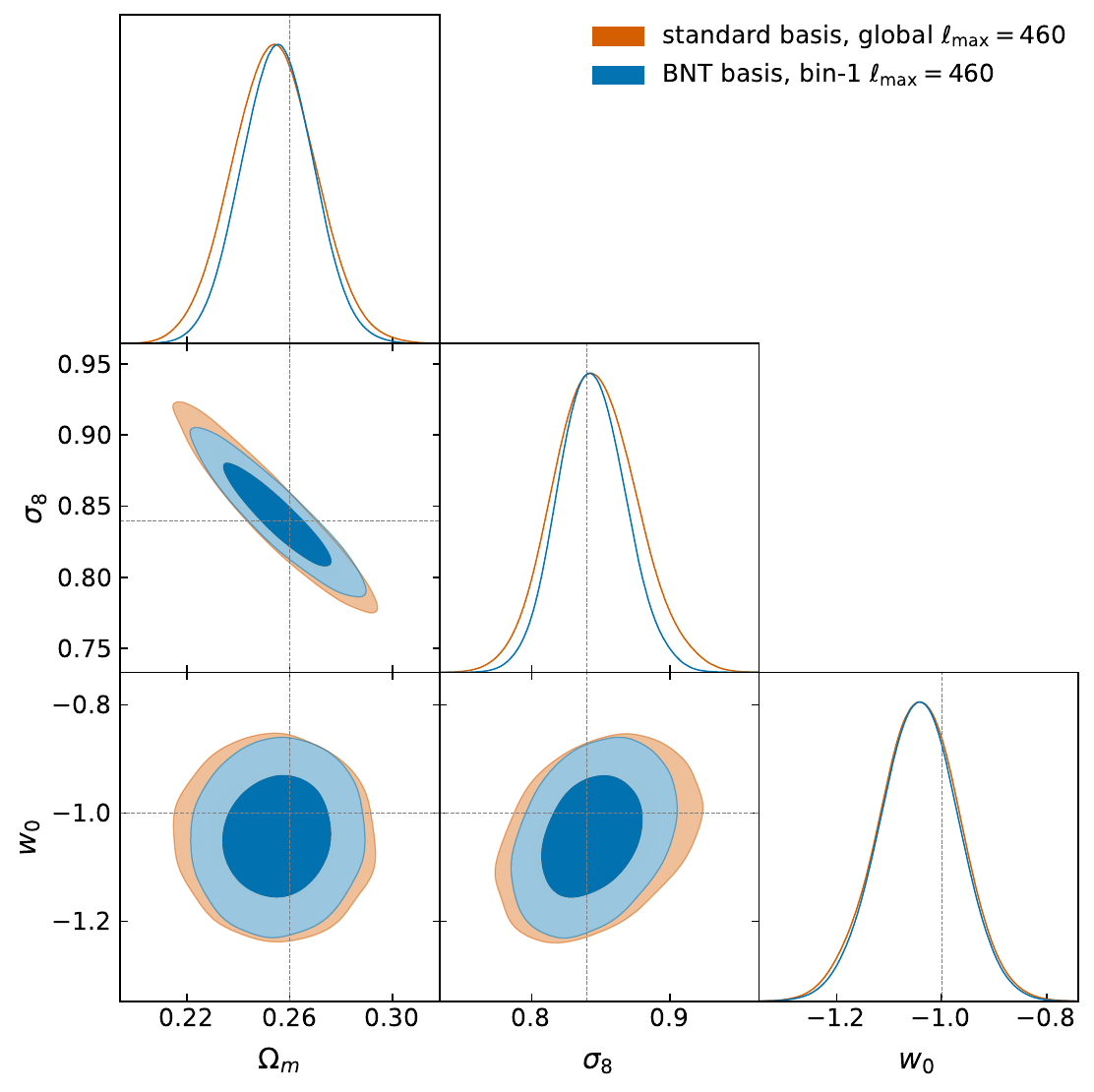}
         \caption{Matched baryon-safe cut ($\ell_{\rm max}=460$).}
         \label{fig:bnt_460}
         \label{fig:bnt_520} % legacy label so old references resolve in the marked-changes diff
     \end{subfigure}

     \caption{
     Comparison of PS posterior constraints in the standard tomographic basis (orange) and the BNT basis (blue) at 14,000 deg$^2$. \textbf{Left:} Constraints at full map resolution ($\ell_{\rm max} = 1024$) without scale cuts. \textbf{Right:} Constraints after applying baryon-safe scale cuts. The standard analysis employs a global cut of $\ell_{\rm max} = 460$ across all bins. The BNT analysis applies the same cut exclusively to the first transformed bin ($\hat{\kappa}_1$), retaining $\ell_{\rm max} = 1024$ for the higher-redshift bins.
     }
     \label{fig:bnt_comparison_combined_ps}
\end{figure*}

We first verify the transform's ability to preserve the information content of the field. In Fig.~\ref{fig:bnt_1024}, we compare the posterior constraints for the PS in the standard and BNT bases at full map resolution ($\ell_{\rm max} = 1024$), without applying any scale cuts. The recovered contours are effectively identical, with the FoM of the two bases agreeing to within a few percent, as expected for an invertible linear transformation. This requires the full data vector, including the cross-spectra between transformed bins, to capture the signal and noise correlations introduced by the mixing matrix.

Obtaining well-calibrated posteriors from the BNT data vector requires one adjustment to the density estimation. The nulling subtractions anticorrelate the transformed bins, so the BNT cross-spectra fluctuate around zero and take negative values, and the cosmological information is spread across many low-amplitude components of the data vector. We therefore prepend a small embedding network to the normalizing flow, which learns a compressed representation of the data vector before density estimation, and we use the same architecture for both bases so that the comparison differs only in the basis. Without this layer, a flow trained on the raw BNT vector recovers accurate marginals but degrades the correlation structure between parameters, and the resulting posteriors understate the information content of the transformed data.

Next, we assess the efficiency of the transform in mitigating baryonic bias. 
In the standard analysis, the broad lensing kernels mix low-redshift baryonic effects into all tomographic bins (see the kernel comparison in Appendix \ref{app:bnt_viz}). Consequently, a global scale cut is required to remove the bias, discarding high-multipole information across the entire survey.
In the BNT basis, however, we confirm that (as was already indicated in Fig. \ref{fig:frac_diff_ps_bnt}) the sensitivity to baryons is effectively localized to the first transformed $z$-bin ($\hat{\kappa}_1$). 
We determined that across all survey areas, applying scale cuts exclusively to this first $z$-bin is sufficient to remove the global baryonic bias. 
We find that the specific $\ell_{\rm max}$ required for this single-bin cut is approximately equal to the corresponding ``global'' scale cut required for the standard analysis.
For the remaining high-redshift bins ($\hat{\kappa}_2, \hat{\kappa}_3, \hat{\kappa}_4$), we can retain scales up to the resolution limit ($\ell_{\rm max} \approx 1024$) without inducing statistically significant tension.

Retaining these high-redshift modes leads to a moderate improvement of the final parameter constraints. At 14,000 deg$^2$ with the matched cut $\ell_{\rm max} = 460$, the bin-specific BNT cut retains 92 of the 120 bandpowers of the data vector, against 50 for the global cut. As shown in Fig.~\ref{fig:bnt_460}, this increases the three-parameter FoM by a factor of $\sim$1.4 over the standard analysis, with the marginal uncertainties on $\Omega_m$ and $\sigma_8$ smaller by $\sim$14\% and $\sim$19\% and essentially unchanged for $w_0$.

\begin{figure*}[t]
     \centering
     % --- First Subfigure ---
     \begin{subfigure}[b]{0.48\textwidth}
         \centering
         \includegraphics[width=\textwidth]{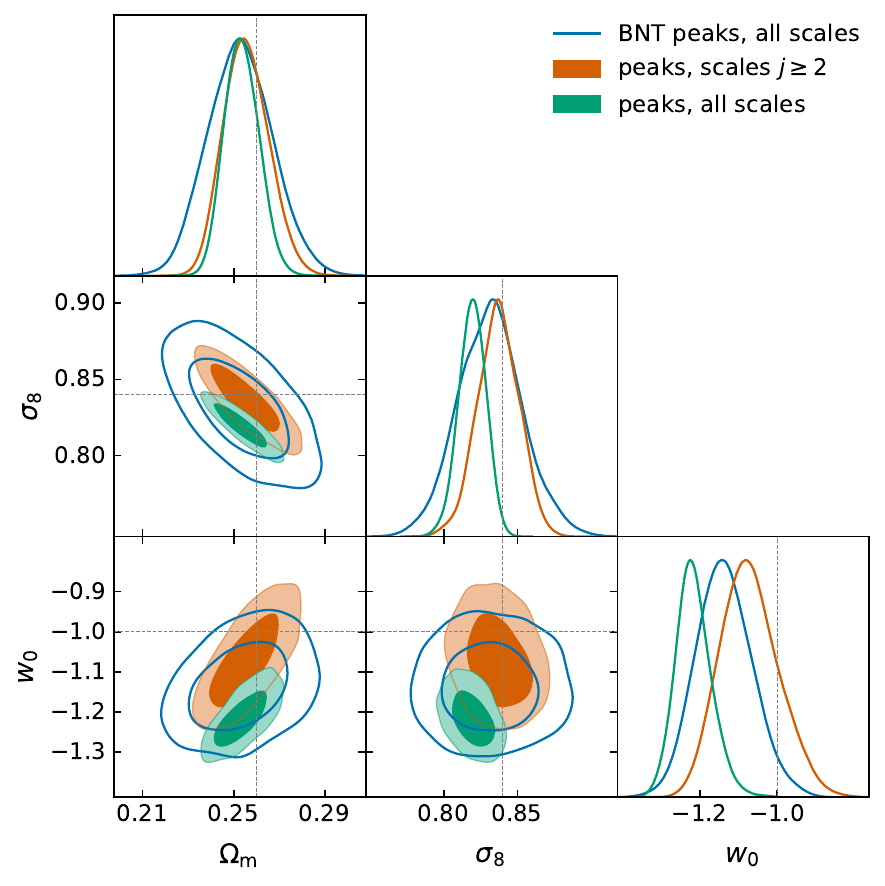}
         \caption{Starlet peak counts}
         \label{fig:bnt_peaks_contours}
     \end{subfigure}
     \hfill
     % --- Second Subfigure ---
     \begin{subfigure}[b]{0.48\textwidth}
         \centering
         \includegraphics[width=\textwidth]{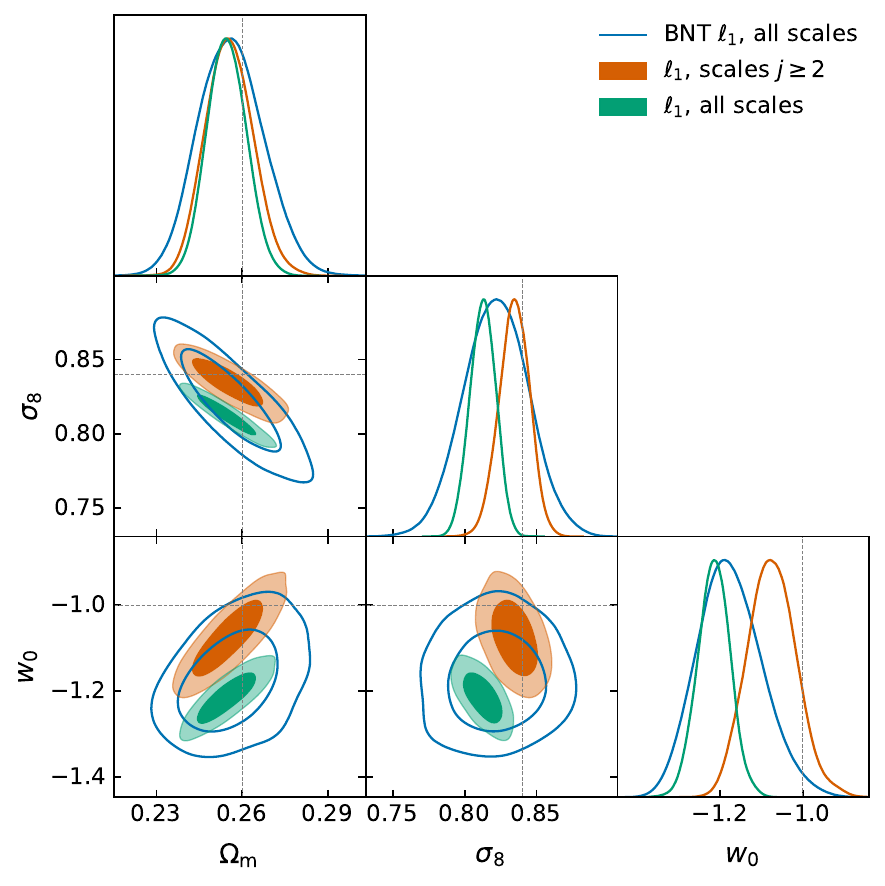}
         \caption{Starlet $\ell_1$-norm}
         \label{fig:bnt_l1_contours}
     \end{subfigure}

     \caption{
     Comparison of posterior constraints for the map-based HOS in the standard basis and the BNT basis at 14,000 deg$^2$, for the baryonified mock data. \textbf{Left:} starlet peak counts. \textbf{Right:} starlet $\ell_1$-norm. Green contours show the standard analysis at full resolution (all wavelet scales), orange contours the standard analysis restricted to safe scales ($j \geq 2$), and the blue outline the BNT analysis at full resolution.
     }
     \label{fig:bnt_comparison_combined}
\end{figure*}

More critically, while the BNT transform performs comparably to the standard analysis for the PS, it faces severe challenges when applied to our map-based HOS. 
Figures \ref{fig:bnt_peaks_contours} and \ref{fig:bnt_l1_contours} demonstrate a drastic inflation of the BNT contours for starlet peak counts and the starlet $\ell_1$-norm compared to the standard analysis, including the case where the latter is restricted by scale cuts.
Even without applying any scale cuts, the BNT contours formally pass our criterion for being unbiased ($n_\sigma < 0.3$); however, this is mainly due to their significantly larger size rather than a true removal of the systematic shift.

Since the BNT transform is a linear, invertible operation, it should in principle be ``lossless'' and conserve the total information content of the field.
We therefore attribute this observed degradation to the noise properties of the transformed field described in Sect. \ref{sec:bnt}, where the subtractions that achieve nulling leave each transformed map with less signal and more noise, and correlate the shape noise between the transformed maps.

As discussed in Appendix \ref{sec:peak_counts}, we work from pre-computed convergence maps and measure the starlet statistics on each tomographic map separately, so our data vector carries only the auto-components. By neglecting the cross-correlations between the BNT maps, we discard information necessary to model this newly correlated noise field. 
While incorporating map-based cross-components, such as generating synthetic ``cross-maps'' in harmonic space \citep{zurcher2022dark, zurcher_towards_2023}, would capture additional cross-bin information and recover some of the lost signal-to-noise, it would not fully resolve the contour inflation.
This is demonstrated by recent Fisher forecasts from the \textit{Euclid} Collaboration \citep{vinciguerra_euclid_2026}, who build tomographic maps for all bin combinations up to quadruplets and retain those that contribute to the figure of merit. Their BNT contours for the map-based HOS are nonetheless inflated relative to the untransformed analysis. This indicates that capturing the complete, uncompressed cross-covariance structure required to make the BNT transform truly ``lossless'' for non-Gaussian statistics is highly non-trivial.

The motivation for nulling extends beyond baryonic feedback. Because lensing projects a range of physical scales onto each angular scale, an angular cut does not correspond to a cut in physical scale. The BNT transform localizes the lensing kernels in distance and restores this correspondence, so that cuts can be placed at a chosen $k_{\rm max}$ \citep{gu2025mitigating}. The same localization makes the field more tractable analytically: predictions for the convergence probability density function (PDF) from large deviation theory rely on nulling to localize the line-of-sight integration \citep{barthelemy_game_2020}, and the nulled frame retains scale-dependent features that projection would otherwise wash out \citep{touzeau_lensing_2026}. These motivations are independent of the summary statistic, and they are why adapting nulling to map-based HOS remains worth pursuing. One approach under investigation filters in the BNT basis and transforms back before the statistics are measured \citep{vinciguerra_euclid_2026}. Another possibility is to pair the transform with denoising, either applied directly or built into the mass-mapping step \citep{tersenov2025impact, leterme2026plug}, which would act on the noise the transform adds.

\section{Conclusions} \label{sec:conclusions}

In this paper, we performed a systematic SBI analysis to quantify the impact of unmodeled baryonic feedback on weak lensing summary statistics (the PS, starlet peak counts, and the starlet $\ell_1$-norm) across a range of survey areas. Using the \texttt{cosmoGRID} simulation suite and neural posterior estimation, we derive the following main conclusions:

\begin{enumerate} 
    \item \textbf{Baryonic bias scales with survey area:} 
    At 2,000 deg$^2$ the parameter shifts induced by unmodeled feedback remain at or below the $1\sigma$ level, and by 5,000 deg$^2$ every statistic exceeds it. At Stage IV-like footprints ($\approx 14,000$ deg$^2$) the shifts reach $2.2\sigma$ for the PS and $3.6\sigma$ for both HOS, and in the full-sky limit they exceed $3.5\sigma$ for the PS and $6\sigma$ for the HOS. These biases are obtained at a maximum multipole of $\ell_{\rm max} = 1024$, well short of the scales that Stage IV analyses aim to exploit. At full map resolution the HOS are the more strongly affected statistics at every area considered.
    
    \item 
    \textbf{Unbiased inference requires discarding most of the small-scale information:} To satisfy our criterion for unbiased inference ($<0.3\sigma$ shift), the largest usable multipole for the PS falls from $\ell = 860$ at 2,000 deg$^2$ to $\ell = 340$ for the full sky, so more than half of the multipole range is removed at Stage IV areas and beyond. For the map-based HOS, excluding the finest wavelet scale ($j=1$) satisfies the same criterion at every area considered. A baryon-safe scale is therefore meaningful only with reference to the survey precision for which it was derived.
    
    \item 
    \textbf{Non-Gaussianity persists on larger scales:} By comparing constraints strictly on these ``baryon-safe'' scales, we find that the starlet $\ell_1$-norm improves the FoM over the PS by a factor of 1.8 at 14,000 deg$^2$ and 2.6 in the full-sky limit, while the starlet peak counts match the PS at Stage IV-like areas and slightly exceed it in the full-sky limit. The HOS are the more baryon-sensitive statistics, and the $\ell_1$-norm nonetheless retains the most cosmological information once each statistic is restricted to its own safe scales. This confirms that significant, cosmologically informative non-Gaussianity is present in the weak lensing field even at intermediate, quasi-linear scales that remain largely unaffected by feedback processes. The HOS contours are also oriented differently from those of the PS in the planes involving $w_0$, so they carry information that is partly complementary to the two-point function.

    \item \textbf{Performance of the BNT transform on the PS:} The transform preserves the information content of the field, provided the full data vector including the cross-spectra between transformed bins is retained. It also localizes the baryonic sensitivity to the lowest transformed redshift bin, so the scale cut can be applied to that bin alone while the higher-redshift bins are kept to the resolution limit. At 14,000 deg$^2$ this improves the figure of merit by a factor of $\sim$1.4 over the standard global cut.
    
    \item \textbf{Performance of the BNT transform on map-based HOS:} 
    The linear mixing required by the transform correlates the shape noise across bins and leaves each transformed map individually signal-poor, washing out the low-amplitude non-Gaussian features that carry much of the signal. The transformed contours formally satisfy our bias criterion, but only because they are substantially larger, and standard tomography with conservative scale cuts yields tighter constraints. For the per-channel statistics used here, the transform therefore offers no net benefit as a baryonic mitigation strategy, consistent with recent Stage-IV forecasts that incorporate cross-tomographic information \citep[e.g.,][]{vinciguerra_euclid_2026}. The transform is linear and invertible, so the information is preserved in the field, whereas our summaries are evaluated on one transformed map at a time. How to construct a summary that captures what the correlations between the maps carry remains an important question for any application of nulling to higher-order statistics.

\end{enumerate}

Our analysis treats baryonic feedback in the most conservative way available. We assume no model for it, and we discard every scale at which it measurably biases the inference. The constraints we obtain are therefore a floor, which any analysis that models the feedback can only improve on. Stage IV analyses will in practice combine scale cuts with explicit modeling of the feedback, drawing on baryonification schemes now developed for higher-order statistics as well as for the power spectrum. They will operate between the floor measured here and the full resolution of the data, where the field is less Gaussian and the gain from higher-order statistics even larger.

\begin{acknowledgements} 
This work was supported by the TITAN ERA Chair project (contract no. 101086741) within the Horizon Europe Framework Program of the European Commission. 
We thank François Lanusse for his conceptual contributions to this work, and Giorgos Valogiannis for useful discussions. We also acknowledge the developers of the \texttt{cosmoGRID} simulation suite for making their data publicly available.
\end{acknowledgements}

\newpage

\bibliographystyle{aa}
\bibliography{new_astro_bib}

\appendix

\section{Weak lensing summary statistics} \label{app:stats}

The robust quantification of the WL signal relies on summary statistics that compress the high-dimensional data (pixelised convergence maps) into a tractable, yet informative, format. We employ three different statistics: the PS, the starlet peak counts, and the starlet $\ell_1$-norm, establishing the classical two-point measure as our baseline before exploring higher-order metrics.

\subsection{Angular power spectrum (PS)}

The PS ($C_{\ell}$) serves as the fundamental baseline measurement in WL analyses. It measures the angular correlations in the convergence field $\kappa$ as a function of angular scale. 
For a gaussian random field with zero mean, the PS encapsulates all available statistical information. While WL mass maps exhibit significant non-gaussianity on small scales due to non-linear gravitational collapse, the PS remains the standard industry metric for extracting cosmological information.

The PS relates the two-dimensional projected correlations to the underlying three-dimensional matter distribution via the Limber approximation:
\begin{equation} 
    C_{ij}(\ell) \propto \int_0^{\chi_{\mathrm{H}}} \frac{\mathrm{d}\chi}{\chi^2} \, W_{i}(\chi) W_{j}(\chi) \, P_{\delta}\left(k = \frac{\ell+1/2}{\chi}, z(\chi)\right), 
\end{equation} 
where $P_\delta (k,z)$ is the 3D matter PS. The term $W_i (\chi)$ represents the lensing efficiency kernel for the $i$-th tomographic redshift bin, defined as: 
\begin{equation}
    W_i(\chi) = \frac{3 \Omega_m H_0^2}{2 c^2} \, (1+z) \, \chi \int_{\chi}^{\chi_{\mathrm{H}}} \mathrm{d}\chi' \, n_i(\chi') \, \frac{\chi' - \chi}{\chi'}, 
\end{equation} 
where $\chi$ is the comoving distance, $\Omega_m$ is the matter density parameter, $H_0$ is the Hubble constant, and $n_i(\chi)$ is the normalized probability distribution of source galaxies in the $i$-th bin.
To exploit the information contained in the growth of structure, we employ lensing tomography, computing the auto-spectra ($i=j$) and cross-spectra ($i \neq j$) for all $z$-bin combinations ($i,j = 1, \dots, n$, and $i\leq j$).

\subsection{Starlet peak counts} \label{sec:peak_counts}

To extract non-Gaussian information from the convergence field, we employ peak count statistics \citep{dietrich_cosmology_2010, kratochvil_probing_2010}, defined as the number of local maxima in the map as a function of their amplitude. Peaks in the weak lensing field are sensitive tracers of the non-linear evolution of structure, corresponding physically to the projection of massive dark matter halos or the alignment of smaller structures along the line of sight \citep{Marian_cosmology_2009}. While high-amplitude peaks typically trace massive galaxy clusters, the abundant population of low-amplitude peaks contains rich cosmological information regarding the background geometry and the growth of structure \citep{dietrich_cosmology_2010}.

\subsubsection{Multiscale decomposition: the starlet transform}

To maximize information extraction across different physical regimes, we perform a multiscale analysis using the isotropic undecimated wavelet transform, commonly known as the ``starlet'' transform \citep{starck:sta06}. 
Unlike Gaussian smoothing, which mixes scales, the starlet transform decomposes the convergence map $\kappa$ into a set of distinct frequency bands--wavelet coefficients. A map of size $N \times N$ is decomposed into a set of wavelet coefficient maps $w_j$ and a coarse-scale residual $c_J$, $\mathcal{W}=\{w_1, \dots, w_{j_{\rm max}}\}, c_J$, such that: 
\begin{equation} 
    \kappa(\boldsymbol{\theta}) = c_J(\boldsymbol{\theta}) + \sum_{j=1}^{J_{\max}} w_j(\boldsymbol{\theta}). 
\end{equation} 
The wavelet functions are compensated (integrating to zero) and have compact support, effectively acting as band-pass filters. The scale index $j$ determines the dyadic resolution, with $j=1$ capturing the finest features (pixel scale) and higher $j$ probing progressively larger angular scales ($\theta \propto 2^j$). 
For our convergence maps at a HEALPix resolution of $N_{\rm side}=512$ (corresponding to a pixel size of $\approx 7\arcmin$), the wavelet at scale $j$ is the difference between smoothings of the map at $2^{j-1}$ and $2^{j}$ times the pixel scale. The nominal sizes of the first four wavelet scales, taken as the geometric means of these kernel pairs, are roughly $10\arcmin$, $20\arcmin$, $40\arcmin$, and $80\arcmin$. The measured multipole coverage of each band is presented in Appendix~\ref{app:starlet_ell}. 
This decomposition decorrelates the information between scales, often resulting in a more diagonal covariance matrix compared to multi-scale Gaussian smoothing \citep{lin_quantifying_2018, ajani_higher_2021}.

\subsubsection{Measurement and signal-to-noise definition}

For each wavelet scale $j$, we identify peaks as pixels in the coefficient map $w_j$ that are strictly larger than their eight nearest neighbors. 
To standardize the measurement against noise variations, we define peaks based on the SNR, $\nu$, of the wavelet coefficients. The SNR map for a given scale $j$ is given by: 
\begin{equation} 
    \nu_j(\mathbf{\theta}) = \frac{w_j(\mathbf{\theta})}{\sigma_j(\mathbf{\theta})}, 
\end{equation} where $\sigma_j(\theta)$ is the noise rms map at scale $j$ \citep{starck:book15}. 
We note here ``SNR'' refers to the local significance or normalized amplitude of the field. This definition preserves the sign of the density fluctuations. 
Since the starlet transform is linear, $\sigma_j$ is computed by convolving the noise variance map of the original data with the square of the starlet filters.
By separating the signal into distinct frequency bands, this statistic allows us to isolate the impact of baryonic feedback—which predominantly affects the finest scales (e.g., $j=1$)—while preserving the clean cosmological signal present in the larger wavelet scales.

\subsubsection{Tomographic cross-components}

In standard weak lensing analyses, cross-tomographic information for higher-order statistics is typically captured by concatenating source galaxy catalogues across different redshift bins prior to map-making. This naturally projects the shared line-of-sight structure into a single combined map, allowing auto-statistics on this joint map to implicitly capture the cross-correlations.  However, because the \texttt{cosmoGRID} suite provides pre-computed convergence maps rather than raw particle or galaxy catalogues, this concatenation approach is not possible for our analysis.

While alternative map-based methods exist to extract cross-tomographic information, such as generating synthetic ``cross-maps'' by multiplying the harmonic space representations of different redshift bins \citep{zurcher2022dark, zurcher_towards_2023}, these techniques only capture a heavily compressed subset of the available cross-information. Specifically, they project the full 2D joint probability distribution of the fields down to a 1D scalar map and are primarily limited to pairwise (two-bin) interactions, missing the complex multi-bin couplings inherent to true higher-order statistics. In this work, to maintain a clean and straightforward baseline for evaluating the scaling of baryonic bias we focus our HOS analysis strictly on the auto-tomographic components, extracting the starlet statistics independently from each $z$-bin.

\subsection{Starlet $\ell_1$-norm}

While peak counts are a powerful probe of high-density regions, they use only the local maxima, effectively discarding the majority of pixels in the map. To capture the full wealth of non-Gaussian information encoded in the convergence field, including the morphology of filaments and the properties of under-dense regions (voids), we employ the starlet $\ell_1$-norm \citep{ajani_starlet_2021}.

This statistic provides a multiscale characterization of the full PDF of the convergence field. Unlike Minkowski functionals or the bispectrum, which can be computationally intensive or difficult to model on masked skies, the starlet $\ell_1$-norm is computationally efficient and robust. It circumvents the ambiguity often associated with defining peaks and voids in noisy weak lensing maps by statistically weighting all pixels rather than selecting discrete minima/maxima.

The statistic is derived from the same starlet decomposition described in Sect. \ref{sec:peak_counts}. Rather than counting discrete features, the starlet $\ell_1$-norm weighs the contribution of all pixels based on their amplitude.
For a given wavelet scale $j$, we define a set of pixels $\mathcal{S}_{j,i}$ whose wavelet coefficients at pixel locations $k$, $w_{j,k}$ fall within a specific SNR bin defined by edges $[B_i, B_{i+1}]$: 
\begin{equation} 
    \mathcal{S}_{j,i} = \left\{ w_{j,k} \Bigm| B_i < \frac{w_{j,k}}{\sigma_j} < B_{i+1} \right\}. 
\end{equation} 
The summary statistic for this bin, $\ell_1^{j,i}$, is the sum of the absolute values of these coefficients (the $\ell_1$-norm of the subset): 
\begin{equation} \label{eq:l1_def} 
    \ell_1^{j,i} = \sum_{w \in \mathcal{S}_{j,i}} |w| = \left| \mathcal{S}_{j,i} \right|_1. 
\end{equation}  
The final data vector consists of the concatenated values of $\ell_1^{j,i}$  across all SNR bins $i$ and wavelet scales $j$.

This calculation functions as a magnitude-weighted histogram. By summing absolute values, the statistic effectively integrates the ``mass'' or ``energy'' of fluctuations in each regime. Since positive wavelet coefficients generally trace overdensities (halos/filaments) and negative coefficients trace underdensities (voids), the binning scheme preserves the physical distinction between these environments while the $\ell_1$ summation amplifies their signal against the Gaussian noise background.

\section{Multipole coverage of the starlet scales} \label{app:starlet_ell}

\begin{figure}[t]
    \centering
    \includegraphics[width=0.48\textwidth]{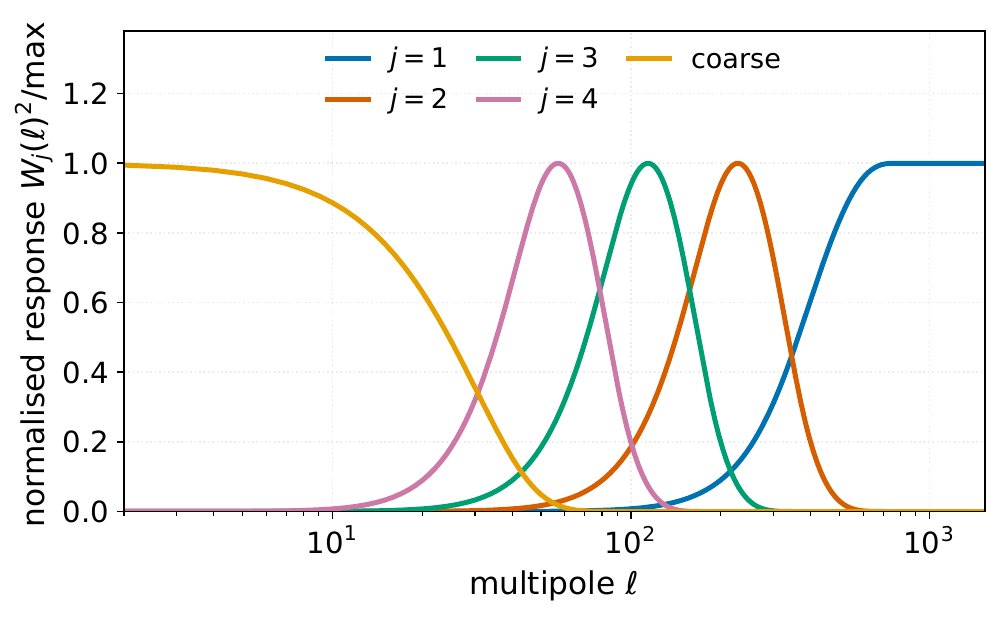}
    \caption{
    \textbf{Starlet band responses.} Peak-normalized squared harmonic response $W_j^2(\ell)$ of the four wavelet scales and the coarse scale of the starlet decomposition at $N_{\rm side}=512$.
    }
    \label{fig:starlet_scale_ell}
\end{figure}

The starlet bands are conventionally labelled by their characteristic angular sizes (Sect.~\ref{sec:hos}). To relate the wavelet bands to the multipoles used for the PS, we measure the harmonic-space response of each band. We apply the spherical starlet decomposition used throughout this work (four wavelet scales and one coarse scale at $N_{\rm side}=512$) to a map containing a single nonzero pixel and compute the ratio of the angular power spectrum of each coefficient map to that of the input map. This yields the squared transfer function $W_j^2(\ell)$ of each band, shown peak-normalized in Fig.~\ref{fig:starlet_scale_ell}. Table~\ref{tab:starlet_bands} lists the peak multipole and the half-power range of each band, defined as the interval over which the response exceeds half its maximum.

Adjacent bands overlap substantially, so the half-power ranges indicate where each band dominates. The finest scale ($j=1$) is resolution-limited. Its response extends to the band limit of the maps ($\ell = 3N_{\rm side}-1 = 1535$), so its coverage is set by the map resolution while the other bands are fixed in multipole. At the opposite end, the coarse scale is confined to $\ell \lesssim 24$ and is the only band that carries the map mean.

On a masked sky, the modes at the lowest multipoles approach the extent of the footprint, the mask couples neighboring multipoles most strongly there, and the mean of the masked map is not an observable. Weak lensing survey analyses therefore start from a minimum multipole well above the fundamental mode \citep[e.g.,][]{asgari_kids1000_2021, dalal_hscy3_2023}, and we follow the same practice, adopting a single multipole range for all survey areas considered in this work. For the wavelet statistics, this exclusion cannot be placed at an arbitrary multipole. The decomposition is dyadic, so the only available cut is the removal of an entire band, and we exclude the coarse scale from the HOS data vectors. The retained wavelet coverage then begins at the half-power edge of the $j=4$ band, $\ell = 36$, and we adopt the first multipole above this edge, $\ell_{\rm min} = 37$, as the lower bound of the PS data vector. All statistics thus begin counting information at the same multipole. Placing this bound instead at $\ell \approx 24$, below which the $j=4$ band retains no appreciable response, leaves our results unchanged.

The responses are measured on the full sphere. On the masked footprints of Sect.~\ref{sec:survey_area_setup}, mode coupling broadens the effective coverage of each band.

\begin{table}
    \centering
    \caption{Measured multipole coverage of the starlet bands at $N_{\rm side}=512$. The half-power range is the interval over which the band response exceeds half its maximum.}
    \label{tab:starlet_bands}
    \begin{tabular}{lccc}
        \hline\hline
        Band & Nominal scale & $\ell$ at peak & Half-power range \\
        \hline
        $j=1$  & $10\arcmin$ & 767 & 364--1535 \\
        $j=2$  & $20\arcmin$ & 228 & 142--336 \\
        $j=3$  & $40\arcmin$ & 114 & 71--168 \\
        $j=4$  & $80\arcmin$ & 57  & 36--84 \\
        coarse & --          & 0   & 0--24 \\
        \hline
    \end{tabular}
\end{table}

\section{Noiseless summary statistics} \label{app:noiseless_stats}

To validate our physical interpretations in Sect. \ref{sec:stats}, and to isolate the intrinsic impact of baryonic feedback from the observational artifacts introduced by galaxy shape noise, we present the fractional differences extracted strictly from noiseless convergence maps.

\begin{figure}[t]
    \centering
    \includegraphics[width=0.45\textwidth]{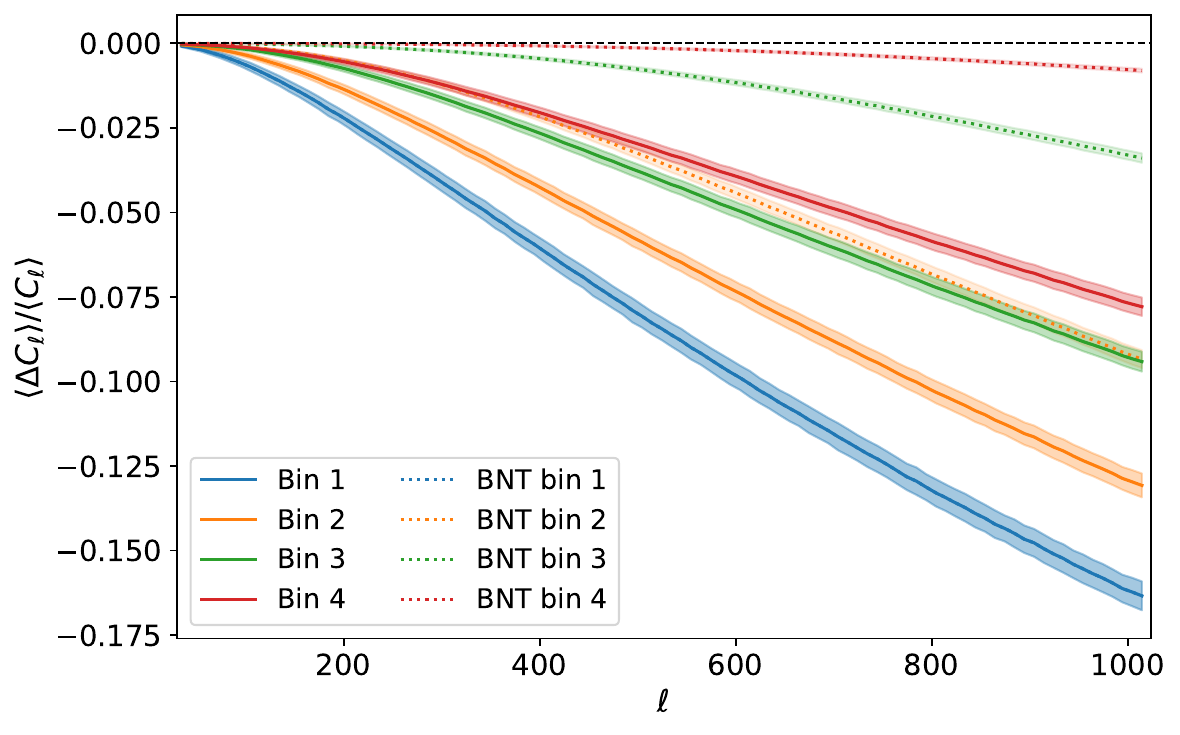}
    \caption{
    Fractional difference in the auto-power spectra between baryonified and DMO simulations, evaluated on strictly noiseless convergence maps. The four tomographic redshift bins are distinguished by color. Solid lines represent the measurements derived in the standard tomographic basis, while dotted lines correspond to the measurements derived in the BNT-transformed basis. Shaded bands indicate the $1\sigma$ scatter across the realizations.
    }
    \label{fig:frac_diff_ps_noiseless}
\end{figure}

Figure \ref{fig:frac_diff_ps_noiseless} illustrates the fractional difference of the auto-power spectra measured on noiseless maps across the four tomographic bins. In the standard tomographic basis (solid lines), the true physical redshift dependence is cleanly resolved. As expected from hydrodynamical models, the lowest redshift bin (Bin 1) is the most heavily impacted by baryonic feedback, diverging from the dark-matter-only (DMO) baseline at lower multipoles and reaching a deeper overall suppression compared to the higher redshift bins. This confirms that the apparent trend observed in the noisy data (where high-redshift bins seemed more biased) is strictly a behavior caused by the shape noise floor.

Furthermore, Figure \ref{fig:frac_diff_ps_noiseless} demonstrates the theoretical efficiency of the BNT transform (dotted lines). By nulling the low-redshift lensing efficiency, the transformation significantly reduces the baryonic impact on the higher redshift bins ($\hat{\kappa}_2, \hat{\kappa}_3, \hat{\kappa}_4$). For these transformed bins, the BNT basis pushes the onset of the baryonic deviation to considerably higher multipoles, effectively expanding the range of ``safe'' scales in the absence of noise.

\begin{figure*}[t]
    \centering
    \includegraphics[width=0.95\textwidth]{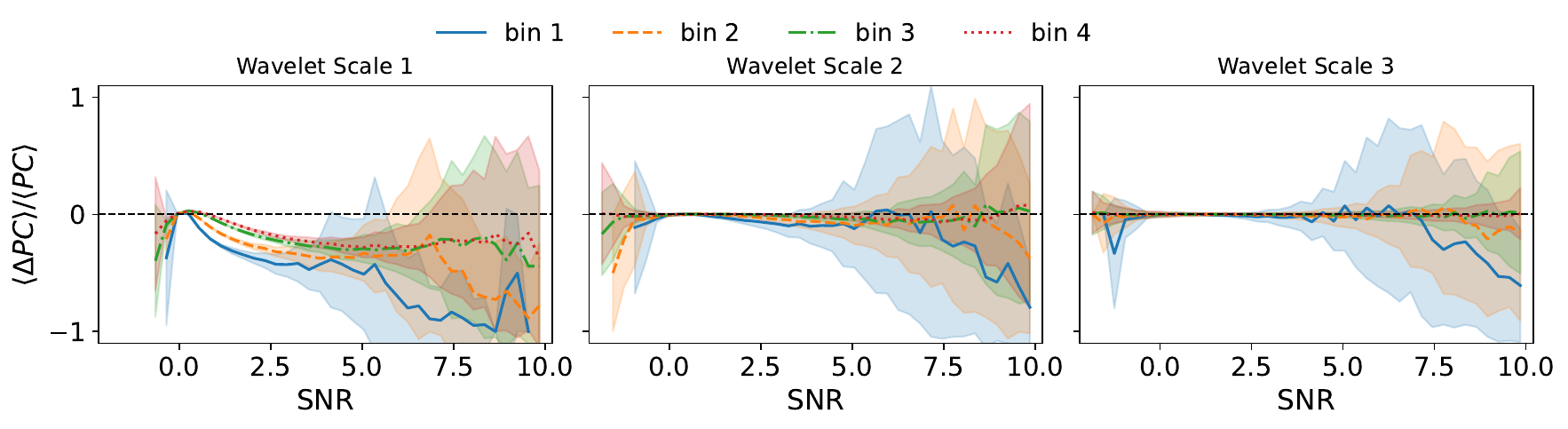}
    \caption{
    Fractional difference in the starlet peak counts between baryonified and DMO simulations, extracted from strictly noiseless convergence maps. The measurements are shown for the first three wavelet scales (left to right panels) as a function of the SNR-equivalent peak height (scaled to match the standard noisy analysis). Solid lines represent the mean fractional difference for each of the four tomographic redshift bins, distinguished by color, with shaded bands indicating the $1\sigma$ scatter.
    }
    \label{fig:frac_diff_pc_noiseless}
\end{figure*}

\begin{figure*}[t]
    \centering
    \includegraphics[width=0.95\textwidth]{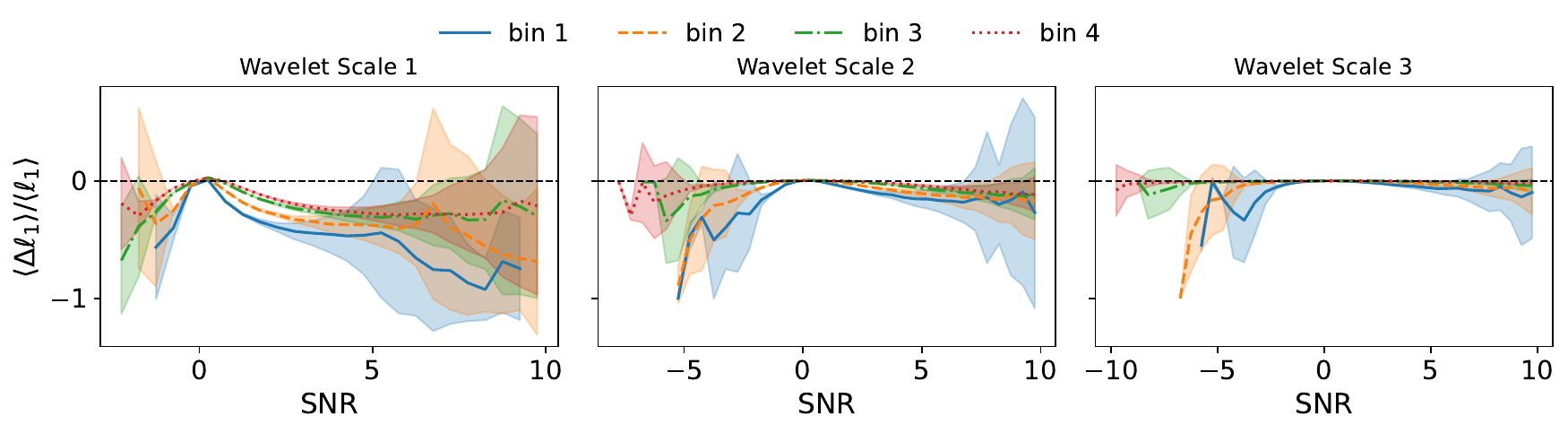}
    \caption{
    Fractional difference in the starlet $\ell_1$-norm between baryonified and DMO simulations, extracted from strictly noiseless convergence maps. The measurements are shown for the first three wavelet scales (left to right panels) as a function of the SNR-equivalent wavelet coefficient magnitude (scaled to match the standard noisy analysis). Solid lines represent the mean fractional difference for each of the four tomographic redshift bins, distinguished by color, with shaded bands indicating the $1\sigma$ scatter.
    }
    \label{fig:frac_diff_l1_noiseless}
\end{figure*}

Figures \ref{fig:frac_diff_pc_noiseless} and \ref{fig:frac_diff_l1_noiseless} present the corresponding noiseless measurements for the starlet peak counts and the starlet $\ell_1$-norm, respectively. 
Consistent with our main analysis, we display the results for the first three wavelet scales ($j=1,2,3$). To facilitate a direct visual comparison with our baseline results, the binning for these noiseless map-based statistics is scaled to match the signal-to-noise ratio (SNR) ranges of the noisy analysis. Because the maps are intrinsically noiseless, these x-axis values represent an ``SNR-equivalent'' peak height or wavelet coefficient magnitude, scaled by the expected shape noise variance of the corresponding noisy maps.
Without the effect of shape noise, both statistics clearly demonstrate that baryonic suppression is minimal near the center of the distribution (intermediate densities) but grows smoothly and continuously toward both the positive and negative high-amplitude tails.

\section{Impact of baryons on starlet peak counts} \label{app:peaks}

As discussed in the main text, the response of starlet peak counts to unmodeled baryonic feedback, and to mitigation strategies like the BNT transform, is qualitatively highly similar to that of the $\ell_1$-norm. Both HOS fundamentally trace the same underlying non-Gaussian density field. To streamline the primary narrative, we present the explicit fractional difference plots for the peak counts in this appendix.

\begin{figure*}[t]
    \centering
    \includegraphics[width=0.95\textwidth]{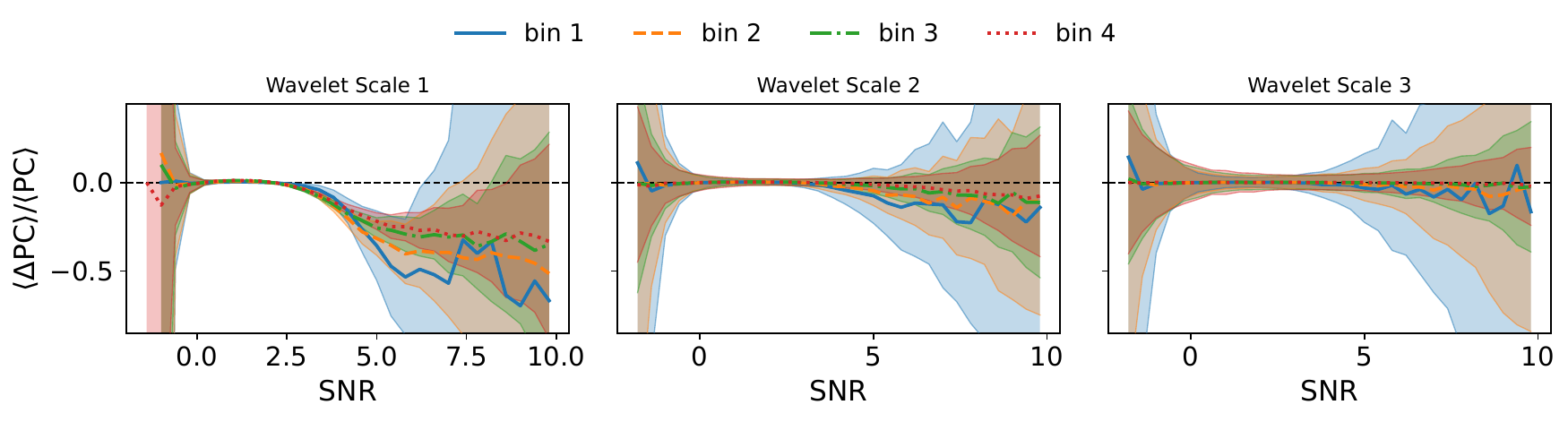}
    \caption{
    Fractional difference in starlet peak counts between baryonified and DMO full-sky maps as a function of SNR ($\nu$), with matched shape-noise realizations between each pair of baryonified and DMO maps. The panels show the first three wavelet scales ($j=1,2,3$, left to right). Lines indicate the mean fractional difference over the fiducial realizations, and shaded bands show the $1\sigma$ statistical uncertainty of the measured peak counts. Bins in which the DMO statistic vanishes are masked.
    } \label{fig:frac_diff_pc}
\end{figure*}

Figure \ref{fig:frac_diff_pc} illustrates the fractional bias induced by the BCM on the peak count statistic across the first three wavelet scales. Consistent with the $\ell_1$-norm analysis, the figure clearly demonstrates the strong scale-dependence of the baryonic effects, with the most severe suppression localized to the finest resolution ($j=1$). The SNR-dependence also reveals the characteristic suppression of high-amplitude peaks ($\nu > 2.5$) associated with gas expulsion from dense halo cores, alongside the inflation of statistical variance at the extreme negative tail ($\nu \in [-1,0]$) where absolute peak counts drop to near-zero.

\begin{figure*}[t]
    \centering
    \includegraphics[width=\textwidth]{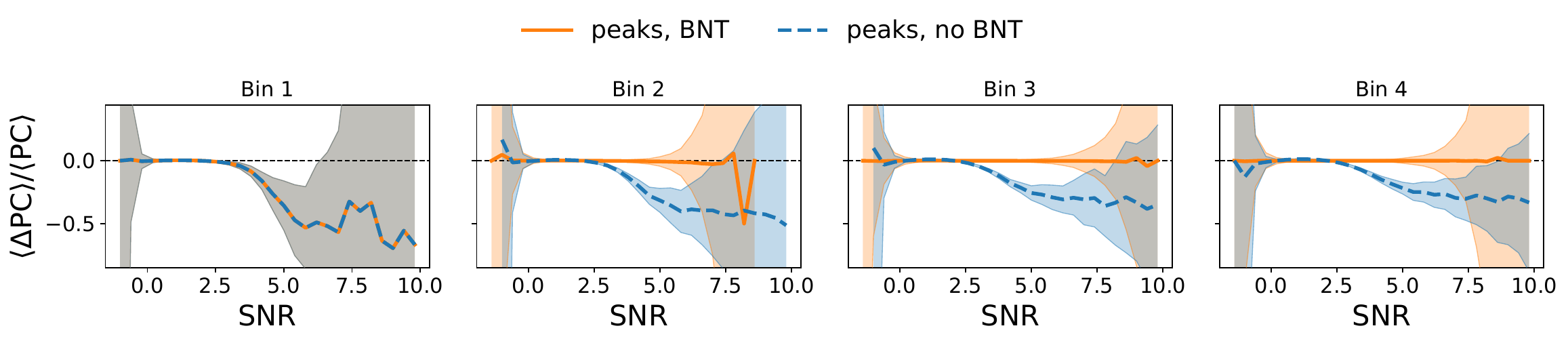}
    \caption{
    Fractional difference in starlet peak counts (first wavelet scale) between baryonified and DMO full-sky maps as a function of SNR ($\nu$) across the four tomographic bins, with shape-noise realizations matched as in Fig.~\ref{fig:frac_diff_pc} and additionally shared between the two bases. Standard tomographic measurements are shown in blue and measurements on the BNT-transformed maps in orange, with shaded bands giving the $1\sigma$ statistical uncertainty of the measured statistic. Bins in which the DMO statistic vanishes are masked. In bin 1 the two bases coincide by construction.
    }
    \label{fig:frac_diff_pc_bnt}
\end{figure*}

Figure \ref{fig:frac_diff_pc_bnt} compares the standard peak counts to the BNT-transformed peak counts for the first, most contaminated wavelet scale. 
Mirroring our findings for the $\ell_1$-norm, the BNT transform is highly effective at mitigating the baryonic bias. While the mean fractional difference does not vanish completely at certain SNR values, the residual shifts are significantly reduced compared to the standard case and remain well within the 1$\sigma$ statistical uncertainty, rendering them statistically insignificant.

\section{Visualizing the BNT transform} \label{app:bnt_viz}

In this appendix, we provide a demonstration of the BNT transform's effect on both the lensing efficiency kernels and the projected mass maps. 

\subsection{Lensing kernels and signal localization}

\begin{figure*}[t]
     \centering
     % --- First Subfigure ---
     \begin{subfigure}[b]{0.48\textwidth}
         \centering
         \includegraphics[width=\textwidth]{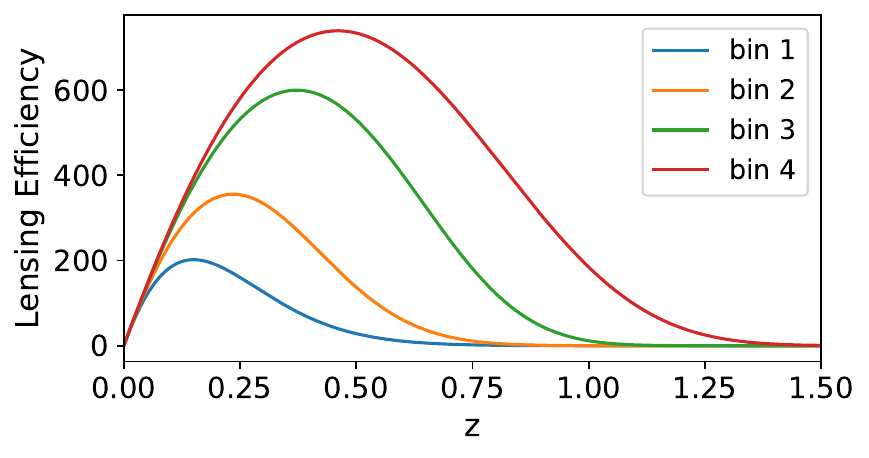}
         \caption{Standard lensing efficiency kernels}
         \label{fig:kernels}
     \end{subfigure}
     \hfill
     % --- Second Subfigure ---
     \begin{subfigure}[b]{0.48\textwidth}
         \centering
         \includegraphics[width=\textwidth]{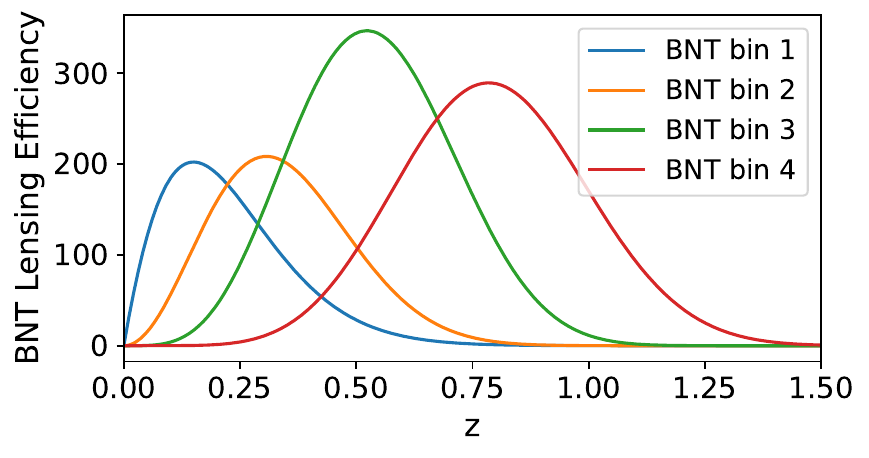}
         \caption{BNT lensing efficiency kernels}
         \label{fig:bnt_kernels}
     \end{subfigure}
     
     \caption{
     \textbf{Lensing efficiency kernels.} \textit{Left panel:} Standard lensing efficiency kernels corresponding to the four tomographic source bins utilized in the analysis. \textit{Right panel:}  Effective lensing efficiency kernels after the application of the BNT transform mixing matrix.
     }
     \label{fig:lensing_kernels}
\end{figure*}

Figure \ref{fig:lensing_kernels} displays the impact of the transformation on the sensitivity of the survey. The left panel shows the standard weak lensing efficiency kernels, $W_i (\chi)$, for the four tomographic bins. As expected, these kernels are broad and heavily overlapping. Due to the cumulative nature of the gravitational lensing effect, higher-redshift bins integrate all structures along the line of sight, effectively containing the signal from lower-redshift bins plus additional contributions from greater distances. This physical mixing necessitates the use of global scale cuts in standard analyses.

In contrast, the BNT-transformed kernels  
$\hat{W}_i (\chi)$ shown in the right panel, demonstrate the ``nulling'' property of the method. By linearly combining the original bins, the transformation cancels out the sensitivity to low-redshift structures. The resulting kernels are significantly more localized, with each BNT bin peaking in a distinct redshift range and suppressing the cumulative signal from the foreground. This localization theoretically allows for redshift-dependent scale cuts.

\subsection{Map-level transformation}

\begin{figure*}[t]
    \centering
    % Top Plot
    \includegraphics[width=0.95\textwidth]{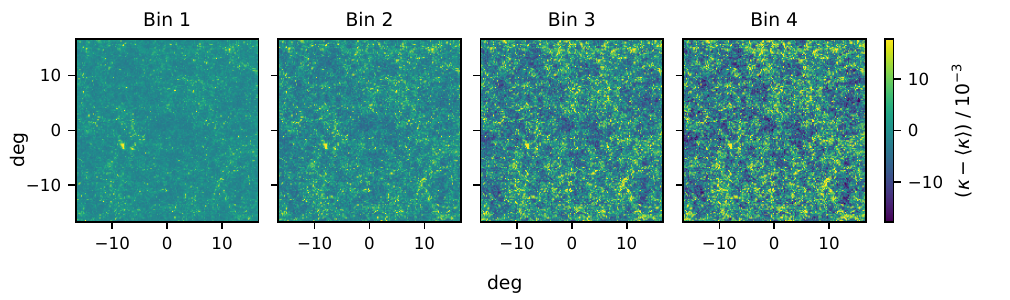}
    
    % \vspace{0.5cm} 
    
    % Bottom Plot
    \includegraphics[width=0.95\textwidth]{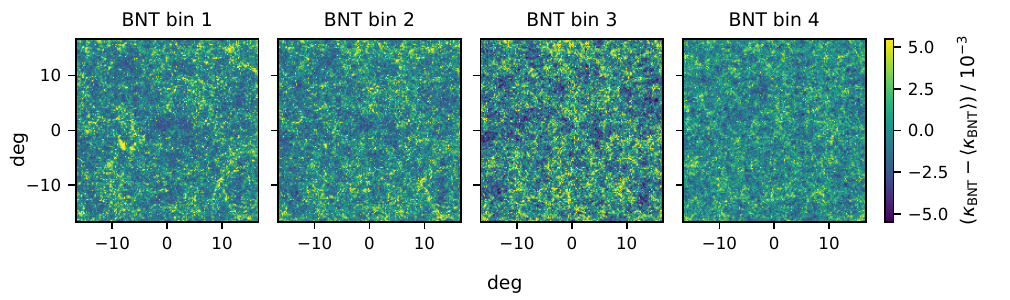}
    
    \caption{
    \textbf{Tomographic maps (noiseless case).} \textit{Top row:} Mean-subtracted convergence maps in the standard tomographic basis for $z$-bins 1 through 4 (increasing in redshift from left to right), for a single noiseless realization. \textit{Bottom row:} The corresponding fields transformed into the BNT basis. Each row shares a single color scale. BNT bin 1 coincides with standard bin 1 by construction.
    }
    \label{fig:noiseless_tomo_maps_stacked_comparison}
\end{figure*}

To visualize the impact on the convergence field, Figure \ref{fig:noiseless_tomo_maps_stacked_comparison} compares the standard and BNT-transformed mass maps for a single realization of a noiseless simulation. In the standard maps (top row), the cumulative nature of lensing is clearly visible: the map for $z$-bin 4 resembles a superposition of the structures seen in $z$-bin 1, with additional structure added from higher redshifts. Conversely, the BNT-transformed maps (bottom row) reveal largely independent structures in each bin. The transform effectively separates the signal originating from different redshift layers, isolating features that were previously blended along the line of sight.

\begin{figure*}[t]
    \centering
    % Top Plot
    \includegraphics[width=0.95\textwidth]{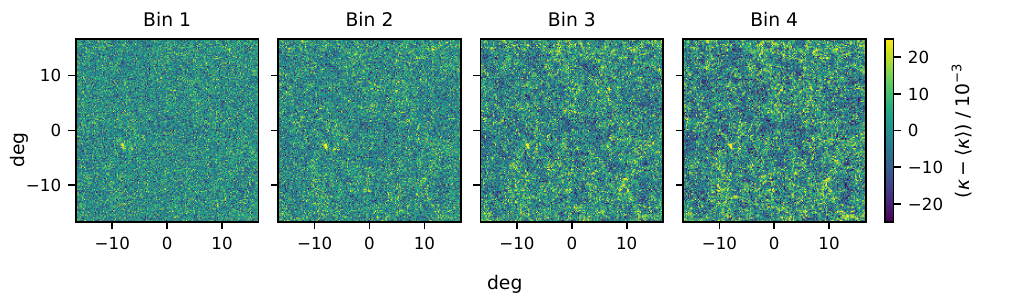}
    
    % \vspace{0.5cm} 
    
    % Bottom Plot
    \includegraphics[width=0.95\textwidth]{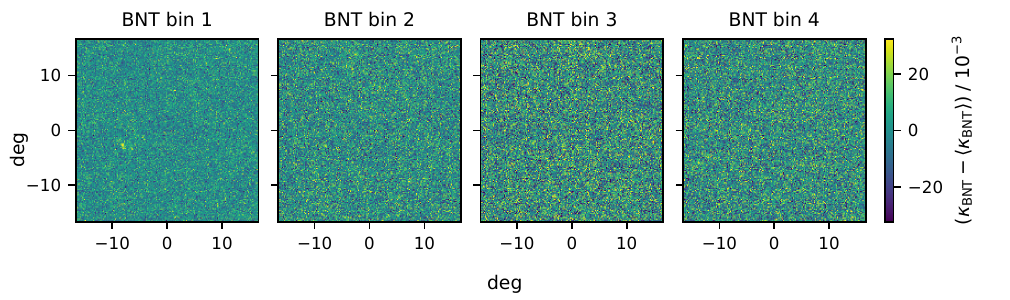}
    
    \caption{
    \textbf{Tomographic maps (noisy case).}  Same as Fig. \ref{fig:noiseless_tomo_maps_stacked_comparison}, but incorporating realistic galaxy shape noise into the original standard convergence maps prior to evaluating the BNT transformation.
    }
    \label{fig:noisy_tomo_maps_stacked_comparison}
\end{figure*}

However, the practical limitation becomes apparent when realistic noise is included. Figure \ref{fig:noisy_tomo_maps_stacked_comparison} shows the same patch with \textit{Euclid}-like shape noise added. 
In the standard basis (top row), the accumulated signal is sufficiently strong that large-scale structures remain visible above the noise even in the higher-redshift bins. In the BNT basis (bottom row), the subtraction process inherently amplifies the noise variance relative to the signal. As a result, the distinct structures visible in the noiseless case are largely obscured by noise. This significant reduction in the per-pixel SNR explains why map-based statistics, such as the starlet $\ell_1$-norm and peak counts, lose constraining power in the transformed basis.

\end{document}